\documentclass[12pt]{article}
\newcommand{\bra}{\langle}
\newcommand{\ket}{\rangle}
\newcommand{\R}{\mbox{\boldmath $ R $}}
\newcommand{\C}{\mbox{\boldmath $ C $}}
\newcommand{\Z}{\mbox{\boldmath $ Z $}}

\newcommand{\vect}[1]{\mbox{\boldmath $ #1 $}}

\makeatletter
\@addtoreset{equation}{section}
\makeatother
\renewcommand{\theequation}{\arabic{section}.\arabic{equation}}

\begin{document} 
\baselineskip 6.9mm 
\begin{flushright}
{\small
The first version: May 31, 2003
\\
Revised on June 7, 2003
\\
hep-th/0306006
\\
KOBE-TH-03-06

}
\end{flushright}
\begin{center}
{\LARGE \bf 
\baselineskip 9mm 
An extension of Fourier analysis for the
$ n $-torus in the magnetic field and its
application to spectral analysis of the
magnetic Laplacian

}
\vspace{8mm}
Makoto Sakamoto\footnote{e-mail: sakamoto@phys.sci.kobe-u.ac.jp},
\quad
Shogo Tanimura\footnote{e-mail: tanimura@kues.kyoto-u.ac.jp}
\vspace*{8mm} \\
{\small \it
${}^1 $Department of Physics, Kobe University, 
Rokkodai, Nada, Kobe 657-8501, Japan
\vspace*{4mm} \\
${}^2 $Department of Engineering Physics and Mechanics, 
Kyoto University \\ Kyoto 606-8501, Japan
}
\vspace{8mm}
\\
{\small Abstract}
\vspace{2mm}
\\
\begin{minipage}[t]{130mm}
\small
\baselineskip 5.6mm 
We solved
the Schr{\"o}dinger equation for a particle
in a uniform magnetic field in the $ n $-dimensional torus.
We obtained a complete set of solutions for a broad class of problems;
the torus $ T^n = \R^n / {\mit\Lambda} $
is defined as a quotient 
of the Euclidean space $ \R^n $ 
by an arbitrary $ n $-dimensional lattice $ {\mit\Lambda} $.
The lattice is not necessary either cubic or rectangular.
The magnetic field is also arbitrary.
However, we restrict ourselves within potential-free problems;
the Schr{\"o}dinger operator is assumed to be
the Laplace operator defined with the covariant derivative.
We defined an algebra that characterizes the symmetry of the Laplacian
and named it the magnetic algebra.
We proved that the space of functions on which the Laplacian acts
is an irreducible representation space of the magnetic algebra.
In this sense the magnetic algebra completely characterizes 
the quantum mechanics in the magnetic torus.
We developed a new method for Fourier analysis for the magnetic torus
and used it to solve the eigenvalue problem of the Laplacian.
All the eigenfunctions are given in explicit forms.
\end{minipage}
\end{center}
\vspace{8mm}
\noindent
{\small
PACS: 
02.40.-k; 
02.40.Gh; 
03.65.-w; 
03.65.Db; 
03.65.Fd 
}

\newpage
\baselineskip 6.2mm 
\section{Introduction}
In this paper we solve the Schr{\"o}dinger equation
for a particle in a uniform magnetic field in an $ n $-dimensional torus.
The problem looks plain at first sight
but actually it turns out to be a hard problem.
Hence we begin this paper by a quick explanation of the problem.
After that we will describe our strategy to solve it.
Subsequently we will briefly mention studies by other people
and describe our motivation of this study.
At the end of Introduction
we will give guides for quick access to main results of this paper.

An $ n $-dimensional torus, or $ n $-torus, is defined as 
$ T^n = \R^n / \Z^n $.
In the coordinate a point 
$ ( t^1, \dots, t^j+1, \dots, t^n ) $
is identified with
$ ( t^1, \dots, t^j, \dots, t^n ) $
for each $ j = 1, \dots, n $.
The eigenvalue problem of the ordinary Laplacian in the torus 
is the equation
\begin{equation}
	- \Delta f 
	=
	- \sum_{j=1}^n 
	\left ( \frac{\partial}{\partial t^j} \right)^2 f
	= 
	\varepsilon f
	\label{ordinary Laplacian}
\end{equation}
with the periodic boundary condition
\begin{equation}
	f ( t^1, \dots, t^j+1, \dots, t^n )
	=
	f ( t^1, \dots, t^j, \dots, t^n ).
	\label{ordinary condition}
\end{equation}
The eigenvalue problem can be immediately solved by Fourier expansion.
A plane-wave function
\begin{equation}
	\chi_k ( t^1, \dots, t^n )
	=
	e^{2 \pi i \sum_{j=1}^n k_j t^j }
	\label{plane-wave}
\end{equation}
with quantized momenta $ k_j \in \Z $ is a solution.
The whole set of eigenfunctions 
$ \{ \chi_k \, | \, (k_1, \dots, $ $ k_n) \in \Z^n \} $ 
constitutes
a complete orthonormal set of the space of periodic functions over the torus.
This is a well-known result.

In this paper we would like to solve an eigenvalue problem of
the magnetic Laplacian;
the magnetic Laplacian is defined by replacing
the partial derivative in the ordinary Laplacian 
by a covariant derivative as
\begin{equation}
	\Delta f =
	\sum_{j,l=1}^n
	g^{jl}
	\left(
		\frac{\partial}{\partial t^j} - 2 \pi i A_j
	\right)
	\left(
		\frac{\partial}{\partial t^l} - 2 \pi i A_l
	\right) f.
	\label{magnetic Laplacian}
\end{equation}
Here $ A_l $ is a component of the $ U(1) $ gauge field
\begin{equation}
	A_{l} 
	= \frac{1}{2} \sum_{j=1}^n \phi_{jl} \, t^j + \alpha_{l} 
\end{equation}
with integers $ \{ \phi_{jl} = - \phi_{lj} \} $ 
and real numbers $ \{ \alpha_j \} $.
The gauge field $ A = \sum_{l=1}^n A_l \, dt^l $ generates
a uniform magnetic field
$ B = dA = (1/2) \sum_{j,l=1}^n \phi_{jl} \, dt^j \wedge dt^l $.
Moreover, 
we would like to consider a general oblique torus
$ T^n = \R^n / {\mit\Lambda} $;
$ {\mit\Lambda} $ is an $ n $-dimensional lattice.
Edges of the unit cell of the lattice
do not necessarily cross at a right angle 
and they do not necessarily have a same length.
Hence we introduce a metric $ g^{jl} $ 
in the definition of the magnetic Laplacian
(\ref{magnetic Laplacian})
to take inclined and stretched or shortened unit cells into account.
The eigenvalue problem of (\ref{magnetic Laplacian}) is accompanied by
the condition
\begin{equation}
	f ( t^1, \dots, t^j+1, \dots, t^n ) 
	=
	e^{\pi i \sum_{k=1}^n \phi_{jk} t^k }
	f ( t^1, \dots, t^j, \dots, t^n ),
	\label{magnetic condition}
\end{equation}
which we call a twisted periodic condition.
Thus the plain problem 
(\ref{ordinary Laplacian}) with
(\ref{ordinary condition})
is generalized to the `magnetic' problem
(\ref{magnetic Laplacian}) with
(\ref{magnetic condition}).
At first glance
it looks rather straightforward
to generalize the problem in this way
but 
it is actually highly nontrivial and difficult
to generalize the solution.

Let us see where the difficulty lies.
In the case of the ordinary Laplacian,
the plane-wave solution (\ref{plane-wave}) is 
a simultaneous eigenfunction of the momentum operators
\begin{equation}
	p_j = -i \frac{\partial}{\partial t^j}
	\qquad
	(j=1, \dots, n)
	\label{ordinary momentum}
\end{equation}
as $ p_j \chi_k = 2 \pi k_j \chi_k $.
The Laplacian can be expressed in terms of the momentum operators
as $ - \Delta = \sum_{j=1}^n (p_j)^2 $
and, of course, it commutes with the momentum operators.
Thus integers 
$ (k_1, \dots, k_n ) $ are `good' quantum numbers.
Then the whole set of simultaneous eigenfunctions $ \{ \chi_k \} $
forms the complete solutions of the Laplacian problem.
This is the way how Fourier analysis works.
However, when we turn to the magnetic Laplacian,
we may seek for a simultaneous eigenfunction of 
`magnetic' momentum operators
\begin{equation}
	p_j = -i 
	\left(
		\frac{\partial}{\partial t^j} - 2 \pi i A_j
	\right)
	\qquad
	(j=1, \dots, n).
	\label{magnetic momentum}
\end{equation}
But such a simultaneous eigenfunction does not exist
because magnetic momenta do not commute with each other 
and instead exhibit commutators
\begin{equation}
	[ \, p_j, p_l \, ] = 2 \pi i \, \phi_{jl}.
	\label{commutator of p}
\end{equation}
The magnetic Laplacian can be still expressed 
in terms of the magnetic momentum operators
as $ - \Delta = \sum_{j=1}^n (p_j)^2 $
but it does not commute with $ p_j $.
Hence, the strategy of ordinary Fourier analysis does not work well
for the magnetic Laplacian.

To solve the problem we developed a new method,
which we call Fourier analysis for the magnetic torus.
This is a main subject of this paper.
Let us describe our strategy:
First, 
we will find 
a group of operators that commute with magnetic momentum operators.
We call the group a magnetic translation group.
Second, 
we enlarge a family of operators
to define an algebra, 
which includes magnetic momenta and magnetic translations as its elements.
We call the algebra a magnetic algebra
and construct its representations.
Third,
we show that 
the space of twisted periodic functions over the torus
is actually an irreducible representation space of the magnetic algebra.
By diagonalizing a maximal commutative subalgebra of the magnetic algebra
we obtain a complete orthonormal set of twisted periodic functions.
This set of orthogonal functions provides a kind of unitary transformation
as the set of plane-wave functions 
provides the Fourier transformation 
which bridges between
the momentum space and the real space.
We note that it is easy to diagonalize the Laplacian in the momentum space.
Finally, we get a whole set of eigenfunctions in the real space
by applying the unitary transformation.
In this procedure
the third step is the hardest part
and is actually accomplished by lengthy cumbersome calculations.
However, the strategy is clear.

We would like to briefly review studies by other people 
on spectral analysis in magnetic field.
Brown \cite{Brown1964} 
first examined the symmetry structure of 
the Schr{\"o}dinger equation for an electron
in a lattice in a uniform magnetic field
and 
found that 
the symmetry is described by a noncommutative discrete translation group.
At the almost same time Zak 
\cite{Zak1964:1}
also found the same symmetry structure
and named the group a magnetic translation group (MTG).
Zak
\cite{Zak1964:2}
immediately built a representation theory of the MTG 
in the three-dimensional lattice.
{}From the viewpoint of functional analysis,
Avron, Herbst, and Simon have been studying
spectral problems of the Schr{\"o}dinger operators in a magnetic field
in a series of papers
\cite{Avron1978, Avron1981, Avron1985}.
Dubrovin and Novikov \cite{Dubrovin1980, Novikov1981}
studied the spectrum of the Pauli operator 
in a two-dimensional lattice with a periodic magnetic field
and intensively analyzed the gap structure above the ground state.
Florek
\cite{Florek2001:1, Florek2001:2}
constructed tensor product representations of the MTG
to analyze a three-particle system in a lattice in a magnetic field.
Kuwabara \cite{Kuwabara1995, Kuwabara1999} has been studying 
quantum-classical correspondence from the viewpoint of spectral geometry.
For example, he \cite{Kuwabara1995} proved that 
if the whole set of level spacings of the quantum spectrum 
is not dense in $ \R $,
every trajectory of the corresponding classical particle 
is a closed orbit.
Arai \cite{Arai} found a quantum plane and quantum group structure
in the quantum system in a singular magnetic field.
Thus we can see that
quantum mechanics in a magnetic field has been an active research area.

Our study on quantum mechanics in magnetic fields
originates from studies of extra-dimension models of the space-time.
In extra-dimension models 
the space-time is assumed to be a base space of a fiber bundle
with a compact fiber or a noncompact fiber.
The history of extra-dimension models is rather old,
but an interest in these models is recently renewed 
as Arkani-Hamed, Dimopoulos, and Dvali
\cite{Arkani-Hamed}
pointed out that the extra-dimension model
may solve the hierarchy problem of high energy physics.
Inspired with extra-dimension models
we \cite{Sakamoto1999:1}
built a model which has a circle $ S^1 $ as a fiber
over an any-dimensional space-times $ \R^{D-1} $.
Then we found that a twisted boundary condition in the $ S^1 $-direction
causes spontaneous breaking of the translational symmetry.
Based on this observation, we \cite{Sakamoto1999:2}
proposed a new mechanism of supersymmetry breaking.
Next we \cite{Tanimura2001, Tanimura2002:1}
built a model which has a two-dimensional sphere $ S^2 $
as a fiber
over the four-dimensional space-times $ \R^4 $.
We solved dynamics in the sphere in a magnetic monopole background
and then found that
the monopole induces spontaneous breaking of 
the rotational symmetry and the $ CP $ symmetry.
We also built a model which has an $ n $-dimensional torus
as a fiber
and tried to analyze dynamics in the torus 
in a background magnetic field.
However, its analysis was not a straightforward task.
Then we studied the symmetry structure of quantum mechanics in the torus
in the magnetic field.
We \cite{Tanimura2002:2}
constructed the MTG in the $ n $-torus and 
classified irreducible representations of the MTG.

Armed with these tools we are now ready to solve the spectral problem
in the $ n $-torus $ T^n = \R^n / {\mit\Lambda} $.
We decide to solve the problem exhaustively;
in our treatment
the dimensions of the torus is taken to be arbitrary,
lengths and angles of edges of the unit cell of $ {\mit\Lambda} $
are arbitrary,
and an arbitrary constant magnetic field is applied to the torus.
Thus we aim to solve the widest class of
quantum mechanics in the $ n $-torus in uniform magnetic fields.

For busy readers here we give guides for quick access to main results.
In Sec. 2 we provide a geometric setting to define the problem.
The problem to be solved 
is the eigenvalue problem of the magnetic Laplacian
(\ref{Laplacian})
with the twist condition
(\ref{twist condition}).
In Sec. 3 we find a family of operators that commute 
with the covariant derivative.
Actually they are composition of 
ordinary displacements and gauge transformations
as shown at 
(\ref{f-v}).
These displacement vectors 
form a restricted family of vectors as shown at
(\ref{magnetic shifts}).
These displacement operators generate the magnetic translation group (MTG),
which is noncommutative as shown at
(\ref{commutator}).
Along (\ref{discrete generators})-(\ref{dim})
we construct irreducible representations of the MTG.
In Sec. 4 we introduce a coordinate system,
which will be revealed to be useful later.
In Sec. 5 we define the magnetic algebra by adding
differential operators (\ref{P_2j-1})-(\ref{P_2m+k})
and
multiplicative operators (\ref{T}) to the MTG.
Then we construct and classify irreducible representations 
of the magnetic algebra.
Sec. 6 is devoted to calculation of simultaneous eigenfunctions
(\ref{chi})
of a maximal commutative subalgebra of the magnetic algebra.
Then we obtain a complete orthonormal set of functions over the magnetic torus,
which provide an extension of Fourier analysis for the magnetic torus.
This is one of main products of this paper.
In Sec. 7 by applying this method we solve the original problem,
the eigenvalue problem of the magnetic Laplacian.
There we obtain a whole set of
eigenfunctions (\ref{solution})
and eigenvalues (\ref{energy eigenvalue}).
These are main results of this paper.

\section{Gauge field in the torus}
Let $ \vect{t} = ( t^1, \dots, t^n ) $ denote a coordinate of
an $ n $-dimensional torus $ T^n = \R^n / \Z^n $.
Namely, a point 
$ ( t^1, \dots, t^j+1, \dots, t^n ) $
is identified with
$ ( t^1, \dots, t^j, \dots, t^n ) $
in $ T^n $.
A uniform magnetic field is generated by the gauge field
\begin{equation}
	A 
	= \sum_{k=1}^n A_{k} \, dt^k
	= \frac{1}{2} \sum_{j,k=1}^n \phi_{jk} \, t^j dt^k 
	+ \sum_{k=1}^n   \alpha_{k} \, dt^k.
	\label{gauge field}
\end{equation}
Here $ \{ \phi_{jk} = - \phi_{kj} \} $ and $ \{ \alpha_j \} $ are 
real constants.
Then the magnetic field is
\begin{equation}
	B = dA 
	= \frac{1}{2} \sum_{j,k=1}^n \phi_{jk} \, dt^j \wedge dt^k.
	\label{magnetic field}
\end{equation}
Therefore,
the number $ \phi_{jk} $ represents magnetic flux which penetrates
the $ (t^j, t^k) $-face of the torus.
We call the array of numbers $ (\phi_{jk}) $ a magnetic flux matrix.

Let us introduce a complex scalar field $ f $ in the torus.
The scalar field couples to the gauge field via the covariant derivative
\begin{equation}
	Df = df - 2 \pi i A f.
	\label{covariant derivative}
\end{equation}
We put the coefficient $ 2 \pi i $ in front of $ A $ for later convenience.
Topology of the torus imposes a boundary condition on the scalar field.
The gauge field itself is not a periodic function on $ \R^n $
but it changes its form as
\begin{equation}
	A ( t^1, \dots, t^j+1, \dots, t^n ) 
	=
	A ( t^1, \dots, t^j, \dots, t^n ) 
	+ \frac{1}{2} \sum_{k=1}^n \phi_{jk} \, dt^k.
	\label{aperidic gauge field}
\end{equation}
Therefore, if we make the gauge transformation
\begin{equation}
	f ( t^1, \dots, t^j+1, \dots, t^n ) 
	=
	e^{\pi i \sum_{k=1}^n \phi_{jk} t^k }
	f ( t^1, \dots, t^j, \dots, t^n ),
	\label{twist condition}
\end{equation}
the covariant derivative (\ref{covariant derivative}) remains covariant as
\begin{equation}
	Df ( t^1, \dots, t^j+1, \dots, t^n ) 
	=
	e^{\pi i \sum_{k=1}^n \phi_{jk} t^k }
	Df ( t^1, \dots, t^j, \dots, t^n ).
\end{equation}
We call the condition (\ref{twist condition}) 
a twisted periodic condition.
There are two ways to bring
a point $ ( t^1, \dots, t^j+1, \dots, t^k+1, \dots, t^n ) $
to $ ( t^1, \dots, t^j, \dots, t^k, \dots, t^n ) $.
The first way is
\begin{eqnarray}
	&&
	f( t^1, \dots, t^j + 1, \dots, t^k + 1, \dots, t^n )
	\nonumber \\
& = &
	e^{\pi i 
		\{ 
		\phi_{jk} 
		+ \sum_{l=1}^n \phi_{jl} t^l 
		\}
	}
	f( t^1, \dots, t^j, \dots, t^k + 1, \dots, t^n )
	\nonumber \\
& = &
	e^{\pi i 
		\{ 
		\phi_{jk} 
		+ \sum_{l=1}^n \phi_{jl} t^l 
		+ \sum_{l=1}^n \phi_{kl} t^l 
		\}
	}
	f( t^1, \dots, t^j, \dots, t^k, \dots, t^n ).
	\label{twist compatibility:1}
\end{eqnarray}
The other way is
\begin{eqnarray}
	&&
	f( t^1, \dots, t^j + 1, \dots, t^k + 1, \dots, t^n )
	\nonumber \\
& = &
	e^{\pi i 
		\{ 
		\phi_{kj} 
		+ \sum_{l=1}^n \phi_{kl} t^l 
		\}
	}
	f( t^1, \dots, t^j + 1, \dots, t^k, \dots, t^n )
	\nonumber \\
& = &
	e^{\pi i 
		\{ 
		\phi_{kj} 
		+ \sum_{l=1}^n \phi_{kl} t^l 
		+ \sum_{l=1}^n \phi_{jl} t^l 
		\}
	}
	f( t^1, \dots, t^j, \dots, t^k, \dots, t^n ).
	\label{twist compatibility:2}
\end{eqnarray}
To make these two expressions coincide we need to have
$$
	e^{\pi i \phi_{jk} } =
	e^{\pi i \phi_{kj} },
$$
namely
\begin{equation}
	e^{\pi i ( \phi_{jk} - \phi_{kj} ) } 
	= e^{2 \pi i \phi_{jk} }
	= 1.
\end{equation}
Therefore, compatibility of the periodic conditions
(\ref{twist compatibility:1}) and
(\ref{twist compatibility:2})
demands that $ \phi_{jk} $ is an integer.
Hence, the magnetic flux through each face of the torus is quantized.
We call the torus where the magnetic field has been introduced
a magnetic torus.

Since two displacements 
$ t^j \mapsto t^j + 1 $ and
$ t^k \mapsto t^k + 1 $ are commutative,
we can write 
the twisted periodic condition (\ref{twist condition})
in a more general form 
\begin{equation}
	f( \vect{t} + \vect{m} )
	=
	e^{ \pi i \sum_{j,k=1}^n \phi_{jk} m^j t^k }
	f( \vect{t} )
	\label{general twist}
\end{equation}
with an arbitrary $  \vect{m} = ( m^1, \dots, m^n ) \in \Z^n $.
An inner product of two twisted periodic functions 
$ f( \vect{t} ) $ and $ g( \vect{t} ) $ is defined by
\begin{equation}
	\bra f | g \ket
	= \int_0^1 dt^1 \cdots \int_0^1 dt^n
	f^* ( \vect{t} ) g( \vect{t} ).
	\label{inner product}
\end{equation}
Equipped with this inner product 
the space of twisted periodic functions becomes a Hilbert space.

To define the Laplacian we need to introduce a metric into the torus.
Let $ {\mit\Lambda} $ be an $ n $-dimensional lattice 
in the Euclidean space $ \R^n $.
We equip the torus $ T^n $ with a Riemannian structure
by identifying $ T^n $ with the quotient space $ \R^n / {\mit\Lambda} $.
Let $ \{ \vect{u}_1, \dots, \vect{u}_n \} $ 
be a set of vectors 
that generates the lattice $ {\mit\Lambda} $.
Their inner products is denoted by
\begin{equation}
	g_{jk} = \bra \vect{u}_j, \vect{u}_k \ket
	\label{metric}
\end{equation}
and its inverse is denoted by $ g^{jk} $.
Then the magnetic Laplacian is defined as
\begin{equation}
	\Delta f =
	\sum_{j,k=1}^n
	g^{jk}
	\left(
		\frac{\partial}{\partial t^j} - 2 \pi i A_j
	\right)
	\left(
		\frac{\partial}{\partial t^k} - 2 \pi i A_k
	\right) f.
	\label{Laplacian}
\end{equation}
It is also referred as the Bochner Laplacian in literature.
The purpose of this paper is to solve the eigenvalue problem 
of the magnetic Laplacian 
accompanied by the twisted periodic condition (\ref{twist condition}).

\section{Magnetic translation group}
Our goal is to find a complete set of eigenvalues and eigenfunctions 
of the magnetic Laplacian (\ref{Laplacian}) as announced above.
A royal road to solving an eigenvalue problem is to detect symmetry.
In this section we determine a group of operators
that commute with the Laplacian
and construct irreducible representations of the group.

The vector space $ \R^n $ acts on the torus as isometries.
However, the gauge field restricts
the admissible class of vectors as seen below.
An arbitrary vector $ \vect{v} \in \R^n $ 
displaces the gauge field (\ref{gauge field}) as
\begin{equation}
	A ( \vect{t} ) \mapsto
	A ( \vect{t} - \vect{v} )
	= A ( \vect{t} ) 
	- d \left( \frac{1}{2} \sum_{j,k=1}^n \phi_{jk} v^j t^k \right).
	\label{A-v}
\end{equation}
If we perform a gauge transformation of the scalar field simultaneously
with the displacement 
\begin{equation}
	f ( \vect{t} ) \mapsto
	f'( \vect{t} )
	= ( U (\vect{v}) f ) ( \vect{t} )
	= 
	e^{\pi i \sum_{j,k=1}^n \phi_{jk} v^j t^k }
	f ( \vect{t} - \vect{v} ),
	\label{f-v}
\end{equation}
then the covariant derivative is changed covariantly 
\begin{equation}
	Df ( \vect{t} ) \mapsto
	Df'( \vect{t} ) 
	= 
	e^{\pi i \sum_{j,k=1}^n \phi_{jk} v^j t^k }
	(Df) ( \vect{t} - \vect{v} ).
\end{equation}
In other words, 
the transformation $ U(\vect{v}) $ commutes with the covariant derivative as
\begin{equation}
	(D U( \vect{v} ) f) ( \vect{t} ) 
	= 
	(U( \vect{v} ) D f) ( \vect{t} ).
	\label{DU=UD}
\end{equation}
Hence it 
commutes with the magnetic Laplacian,
which is defined in terms of the covariant derivative.
The operator $ U(\vect{v}) $ is unitary with respect to the inner product
(\ref{inner product}).

The displaced function (\ref{f-v}) also 
must satisfy the twisted periodic condition.
If the original function $ f $ satisfies the condition (\ref{general twist}),
the displaced function changes its form as
\begin{equation}
	f'( \vect{t} + \vect{m} )
	=
	e^{2\pi i \sum_{j,k=1}^n \phi_{jk} v^j m^k }
	e^{ \pi i \sum_{j,k=1}^n \phi_{jk} m^j t^k }
	f'( \vect{t} )
	\label{twist?}
\end{equation}
for $ \vect{m} \in \Z^n $.
Thus the displaced function 
satisfies the condition (\ref{general twist})
if and only if
\begin{equation}
	\sum_{j,k=1}^n \phi_{jk} v^j m^k 
\end{equation}
is an integer for an arbitrary $ \vect{m} \in \Z^n $.
In other words,
\begin{equation}
	\sum_{j=1}^n \phi_{jk} v^j 
	\qquad (k=1, \dots, n)
	\label{discrete}
\end{equation}
must be an integer.
We call such a restricted vector $ \vect{v} $ a magnetic shift.
The set of magnetic shifts forms an Abelian group
\begin{equation}
	{\mit{\Omega}}^n =
	\{
		\vect{v} \in \R^n \, | \,
		\phi \vect{v} \in \Z^n
	\}.
	\label{magnetic shifts}
\end{equation}
There is a sequence of Abelian subgroups
$ \Z^n \subset {\mit{\Omega}}^n \subset \R^n $.
In particular, an integer vector $ \vect{m} \in \Z^n $ induces
a displacement 
$$
	( U (\vect{m}) f ) ( \vect{t} )
	= 
	e^{\pi i \sum_{j,k=1}^n \phi_{jk} m^j t^k }
	f ( \vect{t} - \vect{m} ).
$$
However, owing to the twisted periodic condition (\ref{general twist}),
this is reduced to the identity transformation
\begin{equation}
	( U (\vect{m}) f ) ( \vect{t} ) = f ( \vect{t} ).
	\label{identity}
\end{equation}
Thus we conclude that the group $ G $
of effective transformations is generated by
\begin{equation}
	\{ U(\vect{v}) \, | \, \vect{v} \in {\mit{\Omega}}^n / \Z^n \}.
	\label{MTG}
\end{equation}
We call the group $ G $ the magnetic translation group (MTG).
A product of transformations is
\begin{equation}
	U (\vect{v}) 
	U (\vect{w}) 
	=
	e^{\pi i \sum_{j,k=1}^n \phi_{jk} v^j w^k }
	U (\vect{v} + \vect{w}).
	\label{product}
\end{equation}
Their commutator is
\begin{equation}
	U (\vect{v}) 
	U (\vect{w}) 
	U (-\vect{v}) 
	U (-\vect{w}) 
	=
	e^{2 \pi i \sum_{j,k=1}^n \phi_{jk} v^j w^k }.
	\label{commutator}
\end{equation}
We can say that
the MTG is a central extension of the Abelian group $ {\mit{\Omega}}^n / \Z^n $
by $ U(1) $.

We can express the MTG in a standard form.
Let $ L (n, \Z) $ 
denote a group of 
the $ n $-dimensional matrices $ \{ S \} $ 
of integers such that $ \det S = \pm 1 $.
The matrix $ S \in L (n, \Z) $ acts on $ \vect{t} \in \R^n $ by
$ \vect{t} \mapsto S \vect{t} $
and this action induces an automorphism of the torus $ T^n = \R^n / \Z^n $.
It also induces a transformation of the magnetic flux matrix as
\begin{equation}
	\phi_{jk} \mapsto 
	\phi'_{jk}
	= \sum_{l,p=1}^n \phi_{lp} \, S^l_{\; j} \, S^p_{\; k}.
	\label{S phi S}
\end{equation}
The Frobenius lemma \cite{Igusa} tells that 
for any integral antisymmetric matrix 
there exists a transformation to bring it into a standard form
\begin{equation}
	( \phi_{jk} )
	=
	\left(
		\begin{array}{cccccccccc}
		 0   & q_1 &      &     &        &      &     &   &        &  \\
		-q_1 & 0   &      &     &        &      &     &   &        &  \\
		     &     &  0   & q_2 &        &      &     &   &        &  \\
		     &     & -q_2 & 0   &        &      &     &   &        &  \\
		     &     &      &     & \ddots &      &     &   &        &  \\
		     &     &      &     &        &  0   & q_m &   &        &  \\
		     &     &      &     &        & -q_m & 0   &   &        &  \\
		     &     &      &     &        &      &     & 0 &        &  \\
		     &     &      &     &        &      &     &   & \ddots &  \\
		     &     &      &     &        &      &     &   &        & 0 
		\end{array}
	\right),
	\label{standard form}
\end{equation}
where $ \{ q_i \} $ are positive integers
and they constitutes a sequence $ q_1 | q_2 | \cdots | q_m $,
which implies $ q_i $ divides $ q_{i+1} $.
For example, we may have a sequence $ 3 | 6 | 12 | 48 $.
Of course, $ 2m \le n $.
The vector subspace of the zero eigenvalue of the matrix $ \phi $ 
has dimensions $ n - 2m $
and it is called null directions.
In the following we suppose that the flux matrix is in the standard form
(\ref{standard form}).

Now we can write the magnetic shifts (\ref{magnetic shifts}) 
in a more explicit form.
Let $ \{ \vect{e}_1, \dots, \vect{e}_n \} $ be the standard basis
of $ \R^n $ 
in the $ (t^1, \dots, t^n ) $-coordinate.
Then any magnetic shift is uniquely expressed as
\begin{equation}
	\vect{v}
	= 
	\sum_{j=1}^m 
	\left(
		  \frac{s_{2j-1}}{q_j} \vect{e}_{2j-1}
		+ \frac{s_{2j}}{q_j} \vect{e}_{2j}
	\right)
	+ \sum_{k=1}^{n-2m} \theta_{2m+k} \vect{e}_{2m+k}
	\label{parametrization of magnetic shift}
\end{equation}
with integers 
$ \{ s_1, \dots, s_{2m} \} $
and real numbers 
$ \{ \theta_{2m+1}, \dots, \theta_n \} $.
Namely, the magnetic shifts are generated by
$ \{
	(1/q_j) \vect{e}_{2j-1},
	(1/q_j) \vect{e}_{2j} \, | \, j=1, \dots, m 
\} $
with integral coefficients and
$ \{
	\vect{e}_{2m+k} \, | \, k=1, \dots, n-2m 
\} $
with real coefficients.
Hence, if the flux matrix $ \phi $ has null directions
$ n - 2m > 0 $,
the MTG has a continuous component.
Otherwise, the MTG is a completely discrete group.

Here we summarize our discussion;
the MTG is generated by the unitary operators
\begin{equation}
	U_j =
	U \left(
		\frac{1}{q_j} \vect{e}_{2j-1}
	\right),
	\quad
	V_j =
	U \left(
		\frac{1}{q_j} \vect{e}_{2j}
	\right)
	\qquad ( j = 1, \dots, m )
	\label{discrete generators}
\end{equation}
and 
\begin{equation}
	W_k ( \theta ) =
	U \left(
		\theta \vect{e}_{2m+k}
	\right)
	\qquad ( k = 1, \dots, n-2m ).
	\label{continuous generators}
\end{equation}
According to 
(\ref{identity}),
(\ref{product}),
(\ref{commutator}) and
(\ref{standard form}),
these generators satisfy the following relations
\begin{eqnarray}
&&	( U_j )^{q_j} = ( V_j )^{q_j} = 1, 
	\label{relation1} \\
&&	U_j V_j U_j^{-1} V_j^{-1} = e^{2 \pi i / q_j}, 
	\label{relation2} \\
&&	W_k ( 1 ) = 1, 
	\label{relation3} \\
&&	W_k ( \theta ) W_k ( \theta' ) = W_k ( \theta + \theta' ) 
	\label{relation4}
\end{eqnarray}
and other trivial commutators.

To solve the eigenvalue problem of the magnetic Laplacian
we need to prepare the whole set of irreducible representations of the MTG.
Let 
$ \{ | r_1, \dots, r_m; d_1, \dots, d_{n-2m} \ket \} $
be elements of a representation space that are labeled by
\begin{eqnarray}
	&& r_j \in \Z / \Z_{q_j} \qquad (j=1, \dots, m), 
	\label{label r}
\\
	&& d_k \in \Z            \qquad (k=1, \dots, n-2m).
	\label{label d}
\end{eqnarray}
Then the generators 
(\ref{discrete generators}) and
(\ref{continuous generators})
are represented by
\begin{eqnarray}
&&	U_j | r_1, \dots, r_j, \dots, r_m; d_1, \dots, d_{n-2m} \ket
	= | r_1, \dots, r_j +1, \dots, r_m; d_1, \dots, d_{n-2m} \ket, 
	\label{represent U} \\
&&	V_j | r_1, \dots, r_j, \dots, r_m; d_1, \dots, d_{n-2m} \ket
	= e^{-2 \pi i r_j/q_j}
	| r_1, \dots, r_j, \dots, r_m; d_1, \dots, d_{n-2m} \ket, 
	\label{represent V} \\
&&	W_k (\theta) | r_1, \dots, r_j, \dots, r_m; d_1, \dots, d_{n-2m} \ket
	= e^{-2 \pi i d_k \theta }
	| r_1, \dots, r_j, \dots, r_m; d_1, \dots, d_{n-2m} \ket.
	\qquad \quad
	\label{represent W}
\end{eqnarray}
Thus 
\begin{equation}
	{\cal H}_d =
	\bigoplus_{r_1 = 0}^{q_1} \cdots
	\bigoplus_{r_m = 0}^{q_m}
	\C | r_1, \dots, r_m; d_1, \dots, d_{n-2m} \ket
\end{equation}
provides an irreducible representation space of the MTG.
Its dimension is
\begin{equation}
	\dim {\cal H}_d = q_1 \times q_2 \times \cdots \times q_m.
	\label{dim}
\end{equation}
The labels $ r_j $ and $ d_k $ in
(\ref{label r}) and
(\ref{label d})
will become good quantum numbers 
for the Laplacian (\ref{Laplacian}).
The dimension (\ref{dim}) will give the degree of degeneracy of 
each eigenvalue.

\section{Diagonalization of the magnetic field strength}
To solve the eigenvalue problem of the magnetic Laplacian
we need to take the metric (\ref{metric}) into account.
Remember that
the basis $ \{ \vect{u}_1, \dots, \vect{u}_n \} $ 
generates the lattice $ {\mit\Lambda} $ 
in the Euclidean space $ \R^n $
and that the torus is isometric to $ \R^n / {\mit\Lambda} $.
Let $ \vect{x} = ( x^1, \dots, x^n ) $ 
be an orthonormal coordinate of $ \R^n $.
It is related to the normalized coordinate 
$ \vect{t} = ( t^1, \dots, t^n ) $ via
\begin{equation}
	\vect{x} = t^1 \vect{u}_1 + \cdots + t^n \vect{u}_n = U \vect{t}.
	\label{x:t}
\end{equation}
In these coordinates the magnetic field (\ref{magnetic field}) is expressed as
\begin{equation}
	B 
	= \frac{1}{2} \sum_{j,k=1}^n B_{jk} \, dx^j \wedge dx^k
	= \frac{1}{2} \sum_{j,k,l,p=1}^n B_{jk} \, u^j_{\;l} \, u^k_{\;p}
	\, dt^l \wedge dt^p
	= \frac{1}{2} \sum_{l,p=1}^n \phi_{lp} \, dt^l \wedge dt^p.
	\label{magnetic field in x}
\end{equation}
The number $ B_{jk} $ is areal density of magnetic flux which penetrates
the $ ( x^j, x^k ) $-plane.
If we perform a coordinate transformation
\begin{equation}
	\vect{x} = R \vect{y}
	\label{x:y}
\end{equation}
by an orthogonal transformation $ R \in O(n,\R) $,
the components of the field strength is transformed as
\begin{equation}
	B 
	= \frac{1}{2} \sum_{j,k=1}^n B_{jk} \, dx^j \wedge dx^k
	= \frac{1}{2} \sum_{j,k,l,p=1}^n B_{jk} \, R^j_{\;l} \, R^k_{\;p}
	\, dy^l \wedge dy^p.
	\label{magnetic field in y}
\end{equation}
By a suitable orthogonal transformation 
the field strength matrix $ ( B_{jk} ) $ can be brought into a standard form
\begin{equation}
	\nu =
	{}^t \!R B \, R
	=
	\left(
		\begin{array}{cccccccccc}
		 0   & \nu_1 &      &     &        &      &     &   &
		         &  \\
		-\nu_1 & 0   &      &     &        &      &     &   &
		        &  \\
		     &     &  0   & \nu_2 &        &      &     &   &
		             &  \\
		     &     & -\nu_2 & 0   &        &      &     &   &
		             &  \\
		     &     &      &     & \ddots &      &     &   &        &  \\
		     &     &      &     &        &  0   & \nu_m &   &
		             &  \\
		     &     &      &     &        & -\nu_m & 0   &   &
		             &  \\
		     &     &      &     &        &      &     & 0 &        &  \\
		     &     &      &     &        &      &     &   & \ddots &  \\
		     &     &      &     &        &      &     &   &        & 0 
		\end{array}
	\right),
	\label{standard form of B}
\end{equation}
where 
$ \{ \nu_j \} $
are positive real numbers.
The $ ( t^1, \dots, t^n ) $-coordinate system 
block-diagonalizes the magnetic flux 
in the form of (\ref{standard form})
while
$ ( y^1, \dots, y^n ) $ 
block-diagonalizes the magnetic field strength 
in the form of (\ref{standard form of B}).
The transformations (\ref{x:t}) and (\ref{x:y}) are combined into 
\begin{equation}
	\vect{y} = {}^t \! R \vect{x} = {}^t \! R U \vect{t} = L \vect{t}.
	\label{y:t}
\end{equation}
Then we obtain relations among the matrices
\begin{equation}
	\phi 
	= {}^t \! U B U 
	= {}^t \! U R \, \nu \, {}^t \! R U 
	= {}^t \! L \nu L.
	\label{magnetic matrix}
\end{equation}
The phase factor in (\ref{f-v}) is rewritten as
\begin{equation}
	\sum_{j,k=1}^n v^j \phi_{jk} t^k
=	\sum_{j,k,l,p=1}^n v^j L^l_{\;j} \nu_{lp} L^p_{\;k} t^k
=	\sum_{j,l,p=1}^n v^j L^l_{\;j} \nu_{lp} \, y^p
=	\sum_{j=1}^n \sum_{l=1}^m
	v^j 
	( L^{2l-1}_{\;j} \nu_{l} y^{2l} - L^{2l}_{\;j} \nu_{l} y^{2l-1} ).
	\label{shift in y}
\end{equation}
The gauge field (\ref{gauge field}) is expressed 
in the $ y $-coordinate as
\begin{equation}
	A 
	= \frac{1}{2} \sum_{j=1}^m \nu_j 
	( y^{2j-1} dy^{2j} - y^{2j} dy^{2j-1} )
	+ \sum_{j,k=1}^n   \alpha_{j} (L^{-1})^j_{\;k} \, dy^k.
	\label{gauge field in y}
\end{equation}
We put 
\begin{equation}
	\beta_k = \sum_{j=1}^n   \alpha_{j} (L^{-1})^j_{\;k}
	\label{beta}
\end{equation}
for later use.

Actually, we can choose a transformation matrix $ L $ 
that has zeros in this pattern
\begin{equation}
	L =
	\left(
	\begin{array}{lll}
	L^{2i-1}_{\;2p-1} & L^{2i-1}_{\;2q} & L^{2i-1}_{\;2m+r} \\
	L^{2j}_{\;2p-1}   & L^{2j}_{\;2q}   & L^{2j}_{\;2m+r}   \\
	L^{2m+k}_{\;2p-1} & L^{2m+k}_{\;2q} & L^{2m+k}_{\;2m+r} 
	\end{array}
	\right)
	=
	\left(
	\begin{array}{lll}
		\ast & 0    & 0   \\
		\ast & \ast & 0   \\
		\ast & \ast & \ast 
	\end{array}
	\right)
	\label{zeros in L}
\end{equation}
with $ i,j,p,q = 1, \dots, m $ and $ k,r = 1, \dots, n-2m $.
The inverse matrix $ L^{-1} $ also has the same pattern of zeros.
This distribution of zeros is proved in the appendix A. 
Then we can rewrite the magnetic shift operators
(\ref{discrete generators})
and (\ref{continuous generators})
with a help of (\ref{shift in y})
in the $ y $-coordinate as
\begin{eqnarray}
	U_j f ( y^i ) 
& = &
	e^{\pi i
		\sum_{l=1}^m
		(1/q_j)
		( L^{2l-1}_{\;2j-1} \nu_{l} y^{2l} 
		- L^{2l}_{\;2j-1} \nu_{l} y^{2l-1} )
	}
	f ( y^i - L^i_{\;2j-1} (1/q_j) ),
	\label{U_j in y} 
\\
	V_j f ( y^i ) 
& = &
	e^{- \pi i
		\sum_{l=1}^m
		(1/q_j)
		L^{2l}_{\;2j} \nu_{l} y^{2l-1} 
	}
	f ( y^i - L^i_{\;2j} (1/q_j) ),
	\label{V_j in y}
\\
	W_k (\theta) f ( y^i ) 
& = &
	f ( y^i - L^i_{\; 2m+k} \theta )
	\label{W_k in y}
\end{eqnarray}
for $ j = 1, \dots, m $ and $ k=1, \dots, n-2m $.
{}From (\ref{zeros in L}) and (\ref{magnetic matrix}) 
we get a formula
\begin{eqnarray}
	\nu_i L^{2i-1}_{\;2j-1} 
& = &	\sum_{l,p=1}^m 
	\nu_l L^{2l-1}_{\;2j-1} \, L^{2l}_{\;2p} \, (L^{-1})^{2p}_{\;2i}
	\nonumber \\
& = &	\sum_{l,p=1}^m 
	\left(
		L^{2l-1}_{\;2j-1} \, \nu_l L^{2l}_{\;2p} -
		L^{2l}_{\;2j-1}   \, \nu_l L^{2l-1}_{\;2p} 
	\right)
	(L^{-1})^{2p}_{\;2i} 
	\nonumber \\
& = &	\sum_{p=1}^m 
	q_j \, \delta_{jp} \, (L^{-1})^{2p}_{\;2i} 
	\nonumber \\
& = &	
	q_j (L^{-1})^{2j}_{\;2i},
	\label{useful formula}
\end{eqnarray}
which will be repeatedly used later.

\section{Magnetic algebra}
In this section 
we introduce new operators 
which act on twisted periodic functions.
These new operators and the operators in the MTG
generate an algebra, which we call a magnetic algebra.
We construct and classify its irreducible representations.
In the next section
we will prove that the space of twisted periodic functions
is actually an irreducible representation of the magnetic algebra.
In this sense, the magnetic algebra completely characterizes 
the quantum mechanics in the magnetic torus.

Now we introduce a family of Hermite operators.
Expanding the covariant derivative (\ref{covariant derivative}) 
in terms of the $ y $-coordinate 
\begin{equation}
	Df = i \sum_{l=1}^n P_l f \, dy^l,
	\label{covariant derivative in y}
\end{equation}
we define differential operators
\begin{eqnarray}
	P_{2j-1} 
& = &	-i \left(
		\frac{\partial}{\partial y^{2j-1}} 
		+ \pi i \nu_j y^{2j} 
		- 2 \pi i \beta_{2j-1}
	\right) 
	\label{P_2j-1} 
\\
	P_{2j} 
& = &	-i \left(
		\frac{\partial}{\partial y^{2j}} 
		- \pi i \nu_j y^{2j-1} 
		- 2 \pi i \beta_{2j}
	\right) 
	\qquad ( j = 1, \dots, m )
	\label{P_2j}
\\
	P_{2m+k} 
& = &	-i \left(
		\frac{\partial}{\partial y^{2m+k}} 
		- 2 \pi i \beta_{2m+k}
	\right) 
	\qquad ( k = 1, \dots, n-2m ).
	\label{P_2m+k}
\end{eqnarray}
These are Hermitian with respect to the inner product
(\ref{inner product}).
Since $ ( y^1, \dots, y^n ) $ is an orthonormal coordinate,
the Laplacian (\ref{Laplacian}) becomes
\begin{equation}
	- \Delta f = \sum_{i=1}^n (P_i)^2 f.
	\label{Laplacian in P}
\end{equation}
Nontrivial commutators among $ P $'s are
\begin{equation}
	[ P_{2j-1}, P_{2j} ] = 2 \pi i \nu_j
	\qquad ( j = 1, \dots, m ).
	\label{commutator of P}
\end{equation}
The other commutators vanish.
We call the operators
$ \{ P_{2j-1}, P_{2j} \} $
transverse momenta
while we call the operators
$ \{ P_{2m+k} \} $
longitudinal momenta.
Since the covariant derivative commutes with 
the magnetic shifts as seen at (\ref{DU=UD}),
the momentum operators 
$ \{ P_i \} $ commute with the shift operators
$ \{ U_j, V_j, W_k \} $,
which are defined at 
(\ref{discrete generators})
and 
(\ref{continuous generators}).

Next we introduce another family of unitary operators.
For this purpose we need an observation;
in the $ t $-coordinate that expresses the magnetic flux matrix 
in the standard form (\ref{standard form}),
$ t^{2m+k} $ $ (k = 1, \dots, n-2m) $ are genuine cyclic coordinates.
That is to say,
the twisted periodic function
(\ref{twist condition})
is periodic with respect to these coordinates as
\begin{equation}
	f ( t^1, \dots, t^{2m+k}+1, \dots, t^n ) 
	=
	f ( t^1, \dots, t^{2m+k}, \dots, t^n )
	\qquad (k = 1, \dots, n-2m)
	\label{periodic condition}
\end{equation}
and the magnetic shift (\ref{f-v})
is reduced to an ordinary continuous shift
\begin{equation}
	( W_k (\theta) f ) ( t^1, \dots, t^{2m+k}, \dots, t^n ) 
	= 
	f ( t^1, \dots, t^{2m+k} - \theta, \dots, t^n ).
	\label{periodic shift}
\end{equation}
Then we define an operator $ T^k $ for each $ k = 1, \dots, n-2m $
which acts on $ f $ by multiplication
\begin{equation}
	( T^k f ) ( \vect{t} )
	= 
	e^{2 \pi i \, t^{2m+k}} f ( \vect{t} )
	= 
	e^{2 \pi i \sum_{i=1}^n (L^{-1})^{2m+k}_{\;i} \, y^i } f ( \vect{t} ).
	\label{T}
\end{equation}
Here we used the inverse $ L^{-1} $ 
of the coordinate transformation (\ref{y:t}).
The operators $ \{ T^k \} $ 
are unitary operators with respect to the inner product
(\ref{inner product}).
They satisfy
\begin{eqnarray}
&&	W_k(\theta) T^k = e^{-2 \pi i \theta} \, T^k W_k(\theta),
	\label{commutator T:W} \\
&&	[ P_i, T^k ] = 2 \pi (L^{-1})^{2m+k}_{\;i} \, T^k
	\qquad ( i=1, \dots, n; \, k = 1, \dots, n-2m )
	\label{commutator P:T}
\end{eqnarray}
and commute with the other generators of the MTG.

Combining all the operators introduced above
we define an algebra $ {\cal A} $ 
with the generators
$ \{ 
	P_i, U_j, V_j, W_k(\theta), T^k \, | 
	\, i = 1, \dots, n; 
	\, j = 1, \dots, m; 
	\, k = 1, \dots, n-2m; 
	\, \theta \in \R 
\} $
and with the relations
(\ref{relation1})-(\ref{relation4}), 
(\ref{commutator of P}),
(\ref{commutator T:W}),
(\ref{commutator P:T})
and other trivial commutators.
We call the algebra $ {\cal A} $ a magnetic algebra.
In the following
we will construct 
all the irreducible representations of the algebra $ {\cal A} $
and classify their unitary equivalence classes.

A subset of generators
$ \{ 
	P_{2j-1}, P_{2m+l}, V_j, W_l(\theta) \, | 
	\, j = 1, \dots, m; 
	\, l = 1, \dots, n-2m; 
	\, \theta \in \R 
\} $
generates a maximal Abelian subalgebra of $ {\cal A} $.
Hence these generators are simultaneously diagonalizable.
Their simultaneous eigenstate $ | k,r,d \ket $
is labeled by
$ r_j \in \Z / \Z_{q_j} $ of (\ref{label r})
and
$ d_l \in \Z $ of (\ref{label d}) with new labels
\begin{equation}
	k_{2j-1}, k_{2m+l} \in \R.
	\label{label k}
\end{equation}
The generators 
$ \{ 
	P_{2j-1}, P_{2m+l} 
\} $
act on these states as
\begin{eqnarray}
&&	P_{2j-1} | k, r, d \ket = 2 \pi k_{2j-1} | k, r, d \ket,
	\label{represent P 2j-1} \\
&&	P_{2m+l} | k, r, d \ket = 2 \pi k_{2m+l} | k, r, d \ket,
	\label{represent P 2m+l} 
\end{eqnarray}
and 
$ \{ 
	V_j, W_l(\theta) 
\} $
act as (\ref{represent V}), (\ref{represent W}), respectively.
The coefficient $ 2 \pi $ was put for later convenience.
Other generators
$ \{ 
	P_{2j}, T^l 
	\, | \, j = 1, \dots, m; \, l = 1, \dots, n-2m
\} $
act on the states as
\begin{eqnarray}
&&	\bra \psi | P_{2j} | k, r, d \ket
	=
	i \nu_j \frac{\partial}{\partial k_{2j-1}} 
	\bra \psi | k, r, d  \ket,
	\label{represent P 2j} \\
&&	T^l | k_i, r, d_k \ket
	=
	e^{2 \pi i
		\{
		\sum_{j=1}^m 
			( k_{2j-1} + \Delta^l k_{2j-1} )
			\Delta^l k_{2j} / \nu_j
		+
		\sum_{p=1}^{n-2m} Z^{lp} d_p
		\}
	}
	| k_i + \Delta^l k_i, r, d_k + \delta_k^{\;l} \ket,
	\qquad
	\label{represent T} 
\end{eqnarray}
with $ \Delta^l k_i = (L^{-1})^{2m+l}_{\;i} $.
We will determine the $ (n-2m) \times (n-2m) $ matrix $ Z^{lp} $ soon later.
Here $ | \psi \ket $ represents an arbitrary state.
The rests  
$ \{ 
	U_j 
	\, | \, j = 1, \dots, m
\} $
act as (\ref{represent U}).
Therefore, in an irreducible representation space
the eigenvalues $ k_{2m+i} $ are linked to $ d_l $ via
\begin{eqnarray}
	k_{2m+i} 
& = &	\sum_{l=1}^{n-2m} d_l (L^{-1})^{2m+l}_{\;2m+i} - \beta_{2m+i}
	\nonumber \\
& = &	\sum_{l=1}^{n-2m} ( d_l - \alpha_{2m+l} )
	(L^{-1})^{2m+l}_{\;2m+i} 
	\qquad (i=1, \dots, n-2m).
	\label{k:d}
\end{eqnarray}
Here the real number $ \beta_{2m+i} $ coincides with 
the one that appeared in (\ref{P_2m+k}).
We also used the relations (\ref{beta}) and (\ref{zeros in L}).

The matrix $ Z^{lp} $ is determined 
by the condition $ T^{l'} T^l = T^l T^{l'} $.
The action of $ T^{l'} T^l $ gives
\begin{eqnarray}
	T^{l'} T^l | k_i, r, d_k \ket
& = &
	e^{2 \pi i
		\{
		\sum_{j=1}^m 
			( k_{2j-1} + \Delta^l k_{2j-1} )
			\Delta^l k_{2j} / \nu_j
		+
		\sum_{p=1}^{n-2m} Z^{lp} d_p
		\}
	}
	\nonumber \\ &&
	e^{2 \pi i
		\{
		\sum_{j=1}^m 
		( k_{2j-1} + \Delta^l k_{2j-1} + \Delta^{l'} k_{2j-1} )
		\Delta^{l'} k_{2j} / \nu_j
		+
		\sum_{p=1}^{n-2m} Z^{l'p} ( d_p + \delta_p^{\;l} )
		\}
	}
	\nonumber \\ &&
	| k_i + \Delta^l k_i + \Delta^{l'} k_i, r, 
	d_k + \delta_k^{\;l} + \delta_k^{\;l'} \ket
	\label{T'T} 
\end{eqnarray}
while the action of $ T^l T^{l'} $ gives
\begin{eqnarray}
	T^{l} T^{l'} | k_i, r, d_k \ket
& = &
	e^{2 \pi i
		\{
		\sum_{j=1}^m 
			( k_{2j-1} + \Delta^{l'} k_{2j-1} )
			\Delta^{l'} k_{2j} / \nu_j
		+
		\sum_{p=1}^{n-2m} Z^{l'p} d_p
		\}
	}
	\nonumber \\ &&
	e^{2 \pi i
		\{
		\sum_{j=1}^m 
		( k_{2j-1} + \Delta^{l'} k_{2j-1} + \Delta^{l} k_{2j-1} )
		\Delta^{l} k_{2j} / \nu_j
		+
		\sum_{p=1}^{n-2m} Z^{lp} ( d_p + \delta_p^{\;l'} )
		\}
	}
	\nonumber \\ &&
	| k_i + \Delta^{l'} k_i + \Delta^{l} k_i, r, 
	d_k + \delta_k^{\;l'} + \delta_k^{\;l} \ket.
	\label{TT'} 
\end{eqnarray}
To give $ T^{l'} T^l = T^l T^{l'} $ the matrix $ Z^{ll'} $ must satisfy
\begin{equation}
	\sum_{j=1}^m 
	( \Delta^{l} k_{2j-1} \Delta^{l'} k_{2j} / \nu_j )
	+
	Z^{l'l} 
	=
	\sum_{j=1}^m 
	( \Delta^{l'} k_{2j-1} \Delta^{l} k_{2j} / \nu_j )
	+
	Z^{ll'}.
	\label{T'T=TT'} 
\end{equation}
A general solution of the above equation is
\begin{equation}
	Z^{ll'} =
	\sum_{j=1}^m 
	( \Delta^{l} k_{2j-1} \Delta^{l'} k_{2j} / \nu_j )
	+ S^{ll'}.
	\label{Z} 
\end{equation}
Here we leave an arbitrary symmetric matrix $ S^{ll'} = S^{l'l} $
yet undetermined.
Actually any choice of $ S^{ll'} $ results 
in an equivalent representation,
and therefore we take $ S^{ll'} = 0 $.

A restricted set of vectors $ \{ |k,r,d \ket \} $ 
that are labeled by the mutually independent parameters
\begin{equation}
	k_{2j-1} \in \R, \,
	r_j      \in \Z / \Z_{q_j}, \,
	d_l \in \Z
	\qquad (j=1, \dots, m; \, l=1, \dots, n-2m)
	\label{independent labels}
\end{equation}
spans a Hilbert space $ {\cal H}_\alpha $
for each fixed value of 
$ ( \alpha_{2m+1}, \alpha_{2m+2}, \dots, \alpha_{n} ) $.
Thus we conclude that
a unitary equivalence class 
of irreducible representations of the algebra $ {\cal A} $ 
has one-to-one correspondence with the parameter
$ ( \alpha_{2m+1}, \alpha_{2m+2}, \dots, \alpha_{n} ) 
\in \R^{n-2m}/\Z^{n-2m} $.

\section{Fourier analysis for the magnetic torus}
Now let us turn to the space of twisted periodic functions.
It is a representation space of the magnetic algebra.
We will calculate the whole family of eigenfunctions 
of the maximal Abelian subalgebra of the magnetic algebra.
These eigenfunctions 
$ \chi_{k,r,d} (t^1, \dots, t^n) = \bra t | k,r,d \ket $
satisfy
\begin{eqnarray}
&&	P_{2j-1} \, \chi_{k,r,d}
= 	-i \left(
		\frac{\partial}{\partial y^{2j-1}} 
		+ \pi i \nu_j y^{2j} 
		- 2 \pi i \beta_{2j-1}
	\right) \chi_{k,r,d}
= 	2 \pi k_{2j-1} \, \chi_{k,r,d} 
	\label{P_2j-1 chi} \\
&&	P_{2j} \, \chi_{k,r,d}
= 	-i \left(
		\frac{\partial}{\partial y^{2j}} 
		- \pi i \nu_j y^{2j-1} 
		- 2 \pi i \beta_{2j}
	\right) \chi_{k,r,d}
=	i \nu_j \frac{\partial}{\partial k_{2j-1}} \chi_{k,r,d}
	\label{P_2j chi} \\
&&	P_{2m+l} \, \chi_{k,r,d}
=	-i \left(
		\frac{\partial}{\partial y^{2m+l}} 
		- 2 \pi i \beta_{2m+l}
	\right) \chi_{k,r,d}
	\nonumber \\ && \qquad \qquad \quad 
=	2 \pi 
	\left(
		\sum_{i=1}^{n-2m} d_i (L^{-1})^{2m+i}_{\;2m+l} 
		- \beta_{2m+l} 
	\right) \chi_{k,r,d}
	\label{P_2m+l chi} \\
&&	U_j \, \chi_{k,r,d}
=	e^{\pi i t^{2j}} \chi_{k,r,d} \left( t^{2j-1} - \frac{1}{q_j} \right)
=	\chi_{k,(r+1),d}
	\label{U_j chi} \\
&&	V_j \, \chi_{k,r,d}
=	e^{-\pi i t^{2j-1}} \chi_{k,r,d} \left( t^{2j} - \frac{1}{q_j} \right)
=	e^{-2 \pi i r_j/q_j} \chi_{k,r,d}
	\label{V_j chi} \\
&&	W_l ( \theta )\, \chi_{k,r,d}
=	\chi_{k,r,d} ( t^{2m+l} - \theta )
=	e^{-2 \pi i d_l \theta } \chi_{k,r,d}
	\label{W_l chi} \\
&&	T^l \, \chi_{k,r,d}
	= 
	e^{2 \pi i \, t^{2m+l}} \chi_{k,r,d}
	= 
	e^{2 \pi i
		\{
		\sum_{j=1}^m 
			( k_{2j-1} + \Delta k_{2j-1} )
			\Delta k_{2j} / \nu_j
		+
		\sum_{p=1}^{n-2m} Z^{lp} d_p
		\}
	}
	\chi_{k + \Delta k,r,(d+1)}
	\qquad
	\label{T^l chi}
\end{eqnarray}
for $ j = 1, \dots, m $;
$ l = 1, \dots, n-2m $,
and $ \theta \in \R $.
Here $ (r+1) $ is abbreviation of $ (r_i +\delta_{ij} ) $
and $ (d+1) $ is abbreviation of $ (d_p + \delta_p^{\;l} ) $.
And $ \Delta k_i = (L^{-1})^{2m+l}_{\;i} $ as given in (\ref{represent T}).
Then 
$ \chi_{k,r,d} $ is a simultaneous eigenfunction of 
$ \{ 
	P_{2j-1}, P_{2m+l}, V_j, W_l(\theta) 
\} $.
In the rest of this section
we will
solve the set of equations (\ref{P_2j-1 chi})-(\ref{T^l chi})
to get the solutions
\begin{eqnarray}
	\chi_{k,r,d} ( y^1, \dots, y^n )
& = &
	c \, 
	e^{-2 \pi i
		\sum_{j=1}^m 
		\sum_{l=1}^{n-2m}
		d_l (L^{-1})^{2m+l}_{\;2j-1} \beta_{2j} / \nu_j 
	}
	\nonumber \\ &&
	e^{- \pi i
		\sum_{j=1}^m \nu_j y^{2j-1} y^{2j} 
	}
	\nonumber \\ &&
	\sum_{\sigma_1,\sigma_2,\dots,\sigma_m = -\infty}^\infty
	\nonumber \\ &&
	e^{ \pi i
		\sum_{i,j,l=1}^m 
		\nu_l
		L^{2l-1}_{\;2i-1}
		L^{2l}_{\;2j-1} 
		( q_i \sigma_i + r_i ) 
		( q_j \sigma_j + r_j ) 
		/ ( q_i q_j )
	} 
	\nonumber \\ &&
	e^{ 2 \pi i
		\sum_{j=1}^m 
		\beta_{2j-1} 
		\{ y^{2j-1} 
		- \sum_{i=1}^m L^{2j-1}_{\;2i-1} (q_i \sigma_i + r_i) / q_i \}
	} 
	\nonumber \\ &&
	e^{ 2 \pi i
		\sum_{j,l=1}^m 
		(q_j \sigma_j + r_j - \gamma_j) (L^{-1})^{2j}_{\; 2l} 
		\{
			y^{2l} 
			- \sum_{i=1}^m L^{2l}_{\;2i-1} 
			( q_i \sigma_i + r_i) / q_i 
		\}
	} 
	\nonumber \\ && 
	e^{ 2 \pi i
		\sum_{k,l=1}^{n-2m}
		d_k ( L^{-1} )^{2m+k}_{\;2m+l} 
		\{
			y^{2m+l} 
			- \sum_{i=1}^m L^{2m+l}_{\;2i-1} 
			( q_i \sigma_i + r_i ) / q_i 
		\}
	}
	\nonumber \\ &&
	e^{ 2 \pi i
		\sum_{j=1}^m 
		k_{2j-1} 
		\{
			y^{2j-1} + \beta_{2j} / \nu_j 
			- \sum_{i=1}^m 
			L^{2j-1}_{\;2i-1} 
			(q_i \sigma_i + r_i - \gamma_i) / q_i
		\}
	},
	\label{chi}
\end{eqnarray}
where $ c $ is a common normalization constant.
The coefficients $ \gamma_j $ are given later at (\ref{gamma}).
The fact that the eigenfunctions are uniquely determined
up to the common coefficient $ c $
implies that
the space of twisted periodic function 
is an irreducible representation space of the magnetic algebra.
Consequently,
the eigenfunctions (\ref{chi}) constitute
a complete orthonormal set of the space of twisted periodic functions
over the torus.
This is one of main results of this paper.
Hence an arbitrary twisted periodic function 
over the torus can be expanded as
\begin{equation}
	f (y^1, \dots, y^n) 
	= \sum_{k,r,d} \lambda_{k,r,d} \, \chi_{k,r,d} (y^1, \dots, y^n)
	\label{CONS}
\end{equation}
with unique coefficients $ \lambda_{k,r,d} $.
Therefore,
the complete set $ \{ \chi_{k,r,d} \} $
provides a new basis for Fourier analysis in the magnetic torus.

In the rest of this section we give detailed lengthy calculations
to prove the above statements.
The reader may skip them to the next section,
where we calculate 
solutions of the eigenvalue problem of the magnetic Laplacian
using the main result (\ref{chi}).
First, a simultaneous solution of 
(\ref{P_2j-1 chi}) and (\ref{P_2m+l chi}) is 
\begin{eqnarray}
	\chi_{k,r,d} ( y^1, y^2, \dots, y^n ) 
& = &
	e^{ 2 \pi i 
	\sum_{j=1}^m 
		\{ 
			(k_{2j-1} + \beta_{2j-1}) y^{2j-1} 
			- (1/2) \nu_j y^{2j-1} y^{2j} 
		\}
	}
	\nonumber \\ && 
	e^{ 2 \pi i 
	\sum_{j,l=1}^{n-2m}
		d_j ( L^{-1} )^{2m+j}_{\;2m+l} \, y^{2m+l} 
	}
	\phi_{k,r,d} ( y^2, y^4, \dots, y^{2m} ),
	\label{eigen:1}
\end{eqnarray}
where $ \phi_{k,r,d} ( y^2, y^4, \dots, y^{2m} ) $ is an arbitrary function
to be specified later.

Next let us turn to the other equation (\ref{V_j chi}).
Using (\ref{V_j in y}) we can rewrite (\ref{V_j chi}) as
\begin{equation}
	e^{- \pi i
		\sum_{l=1}^m
		(1/q_j)
		L^{2l}_{\;2j} \nu_{l} y^{2l-1} 
	}
	\chi_{k,r,d} ( y^i - L^i_{\;2j} /q_j )
=	e^{-2 \pi i r_j/q_j} \chi_{k,r,d} ( y^i ).
	\label{V chi rewritten}
\end{equation}
As discussed at (\ref{zeros in L}) we have taken the matrix $ L $ such that
$ L^{2l-1}_{\;2j} = 0 $.
Therefore, when (\ref{eigen:1}) is substituted,
the LHS of (\ref{V chi rewritten}) becomes
\begin{eqnarray}
&&	e^{- \pi i
		\sum_{l=1}^m
		(1/q_j)
		L^{2l}_{\;2j} \nu_{l} y^{2l-1} 
	}
	\chi_{k,r,d} ( y^i - L^i_{\;2j} /q_j )
	\nonumber \\
& = &
	e^{ 2 \pi i 
	\sum_{l=1}^m 
		\{ 
			(k_{2l-1} + \beta_{2l-1}) y^{2l-1} 
			- (1/2) \nu_l y^{2l-1} y^{2l} 
		\}
	}
	\, e^{ 2 \pi i 
	\sum_{p,l=1}^{n-2m}
		d_p ( L^{-1} )^{2m+p}_{\;2m+l} \, y^{2m+l} 
	}
	\nonumber \\ && 
	e^{ - 2 \pi i 
	\sum_{p,l=1}^{n-2m}
		d_p ( L^{-1} )^{2m+p}_{\;2m+l} L^{2m+l}_{\; 2j} /q_j 
	}
	\phi_{k,r,d} ( y^{2i} - L^{2i}_{\; 2j} /q_j ).
	\label{V chi LHS}
\end{eqnarray}
Hence, if we put 
\begin{equation}
	\gamma_j =
	\sum_{p,l=1}^{n-2m}
	d_p ( L^{-1} )^{2m+p}_{\;2m+l} L^{2m+l}_{\; 2j},
	\label{gamma}
\end{equation}
(\ref{V chi rewritten}) implies that
\begin{equation}
	e^{- 2 \pi i \gamma_j/q_j }
	\phi_{k,r,d} ( y^{2i} - L^{2i}_{\; 2j} /q_j )
	=
	e^{-2 \pi i r_j / q_j} 
	\phi_{k,r,d} ( y^{2i} ).
	\label{quasi periodicity}
\end{equation}
If we introduce another coordinate system $ (z^1, z^2, \dots, z^m) $
which is related to $ (y^2, y^4, \dots, y^{2m} ) $ via
\begin{equation}
	y^{2i} = \sum_{j=1}^m L^{2i}_{\;2j} \, (z^j/q_j),
	\label{y:z}
\end{equation}
then (\ref{quasi periodicity}) is rewritten as
\begin{equation}
	\phi_{k,r,d} ( z^1, \dots, z^j - 1, \dots, z^m )
	=
	e^{2 \pi i ( \gamma_j - r_j ) / q_j } \,
	\phi_{k,r,d} ( z^1, \dots, z^j, \dots, z^m ).
	\label{quasi periodicity in z}
\end{equation}
Moreover, if we put
\begin{equation}
	\psi_{k,r,d} ( z^1, \dots, z^m )
	=
	e^{2 \pi i \sum_{j=1}^m ( \gamma_j - r_j ) z^j / q_j } \,
	\phi_{k,r,d} ( z^1, \dots, z^m ),
	\label{psi in z}
\end{equation}
then (\ref{quasi periodicity in z}) implies that
\begin{equation}
	\psi_{k,r,d} ( z^1, \dots, z^j - 1, \dots, z^m )
	=
	\psi_{k,r,d} ( z^1, \dots, z^j, \dots, z^m ).
\end{equation}
Hence $ \psi_{k,r,d} $ is a periodic function with the period $ 1 $
and can be expanded in a Fourier series
\begin{equation}
	\psi_{k,r,d} ( z^1, \dots, z^j, \dots, z^m )
	=
	\sum_{\sigma_1,\sigma_2,\dots,\sigma_m = -\infty}^\infty
	c_{k,r,d,\sigma} \, e^{2 \pi i \sum_{j=1}^m \sigma_j z^j }.
\end{equation}
Note that the inverse transformation of (\ref{y:z}) is given by
\begin{equation}
	(z^j /q_j)
	= \sum_{i=1}^m (L^{-1})^{2j}_{\;2i} \, y^{2i}.
	\label{z:y}
\end{equation}
Combining the above equations
we can write down the eigenfunction (\ref{eigen:1}) 
in a more specific form as
\begin{eqnarray}
	\chi_{k,r,d} ( y^1, y^2, \dots, y^n ) 
& = &
	e^{ 2 \pi i
	\sum_{j=1}^m 
		\{ 
			(k_{2j-1} + \beta_{2j-1}) y^{2j-1} 
			- (1/2) \nu_j y^{2j-1} y^{2j} 
		\}
	}
	\nonumber \\ && 
	e^{ 2 \pi i
		\sum_{j,l=1}^{n-2m}
		d_j ( L^{-1} )^{2m+j}_{\;2m+l} \, y^{2m+l} 
	}
	\nonumber \\ &&
	\sum_{\sigma_1,\sigma_2,\dots,\sigma_m = -\infty}^\infty
	c_{k,r,d,\sigma} \, 
	e^{ 2 \pi i
		\sum_{j,l=1}^m 
		(q_j \sigma_j + r_j - \gamma_j) (L^{-1})^{2j}_{\; 2l} \, y^{2l} 
	}.
	\label{eigen:2}
\end{eqnarray}
Moreover, referring to (\ref{zeros in L}) and (\ref{W_k in y}),
we can see that (\ref{eigen:2}) satisfies (\ref{W_l chi}).
Thus we have seen that
$ \chi_{k,r,d} $ is a simultaneous eigenfunction of 
$ \{ 
	P_{2j-1}, P_{2m+l}, V_j, W_l(\theta) 
\} $ as announced above.

The remaining task is to solve
(\ref{P_2j chi}),
(\ref{U_j chi}), and
(\ref{T^l chi}).
Let us begin with (\ref{P_2j chi}).
The LHS of (\ref{P_2j chi}) is 
\begin{eqnarray*}
&&	-i \left(
		\frac{\partial}{\partial y^{2j}} 
		- \pi i \nu_j y^{2j-1} 
		- 2 \pi i \beta_{2j}
	\right) \chi_{k,r,d}
	\nonumber \\
& = &
	\sum_{\sigma_1,\sigma_2,\dots,\sigma_m = -\infty}^\infty
	2 \pi \left(
		- \frac{1}{2} \nu_j y^{2j-1} 
		+ \sum_{i=1}^m 
		(q_i \sigma_i + r_i - \gamma_i) (L^{-1})^{2i}_{\; 2j} 
		- \frac{1}{2} \nu_j y^{2j-1} 
		- \beta_{2j}
	\right) 
	\nonumber \\ && 
	e^{ 2 \pi i
	\sum_{i=1}^m 
		\{ 
			(k_{2i-1} + \beta_{2i-1}) y^{2i-1} 
			- (1/2) \nu_i y^{2i-1} y^{2i} 
		\}
	}
	e^{ 2 \pi i
		\sum_{i,l=1}^{n-2m}
		d_i ( L^{-1} )^{2m+i}_{\;2m+l} \, y^{2m+l} 
	}
	\nonumber \\ &&
	c_{k,r,d,\sigma} \, 
	e^{ 2 \pi i
		\sum_{i,l=1}^m 
		(q_i \sigma_i + r_i - \gamma_i) (L^{-1})^{2i}_{\; 2l} \, y^{2l} 
	}
\end{eqnarray*}
On the other hand, the RHS of (\ref{P_2j chi}) is 
\begin{eqnarray*}
	i \nu_j \frac{\partial}{\partial k_{2j-1}} \chi_{k,r,d}
& = &
	i \nu_j \,
	e^{ 2 \pi i
	\sum_{i=1}^m 
		\{ 
			(k_{2i-1} + \beta_{2i-1}) y^{2i-1} 
			- (1/2) \nu_i y^{2i-1} y^{2i} 
		\}
	}
	e^{ 2 \pi i
		\sum_{i,l=1}^{n-2m}
		d_i ( L^{-1} )^{2m+i}_{\;2m+l} \, y^{2m+l} 
	}
	\nonumber \\ &&
	\sum_{\sigma_1,\sigma_2,\dots,\sigma_m = -\infty}^\infty
	\left(
		\frac{\partial c_{k,r,d,\sigma} }{\partial k_{2j-1}}
		+ 2 \pi i y^{2j-1} \, c_{k,r,d,\sigma}
	\right)
	e^{ 2 \pi i
		\sum_{i,l=1}^m 
		(q_i \sigma_i + r_i - \gamma_i) (L^{-1})^{2i}_{\; 2l} \, y^{2l} 
	}.
\end{eqnarray*}
Therefore, we have an equation
\begin{equation}
	i \nu_j \,
	\frac{\partial c_{k,r,d,\sigma} }{\partial k_{2j-1}}
	=
	2 \pi \left(
		\sum_{i=1}^m 
		(q_i \sigma_i + r_i - \gamma_i) (L^{-1})^{2i}_{\; 2j} 
		- \beta_{2j}
	\right) 
	c_{k,r,d,\sigma}
\end{equation}
and get its solution
\begin{equation}
	c_{k,r,d,\sigma}
	=
	c_{0,r,d,\sigma} \,
	e^{- 2 \pi i 
	\{
		\sum_{i,j=1}^m 
		(q_i \sigma_i + r_i - \gamma_i) (L^{-1})^{2i}_{\; 2j} 
		\, k_{2j-1} / \nu_j
		- 
		\sum_{j=1}^m \beta_{2j} k_{2j-1} / \nu_j
	\}
	}.
\end{equation}
Thus (\ref{eigen:2}) becomes
\begin{eqnarray}
	\chi_{k,r,d} ( y^1, y^2, \dots, y^n ) 
& = &
	e^{ 2 \pi i
	\sum_{j=1}^m 
		\{ 
			( k_{2j-1} + \beta_{2j-1} ) y^{2j-1} 
			+ k_{2j-1} \beta_{2j} / \nu_j 
			- (1/2) \nu_j y^{2j-1} y^{2j} 
		\}
	}
	\nonumber \\ && 
	e^{ 2 \pi i
		\sum_{j,l=1}^{n-2m}
		d_j ( L^{-1} )^{2m+j}_{\;2m+l} \, y^{2m+l} 
	}
	\nonumber \\ &&
	\sum_{\sigma_1,\sigma_2,\dots,\sigma_m = -\infty}^\infty
	c_{0,r,d,\sigma} \,
	e^{ 2 \pi i
		\sum_{j,l=1}^m 
		(q_j \sigma_j + r_j - \gamma_j) (L^{-1})^{2j}_{\; 2l} 
		( y^{2l} - k_{2l-1} / \nu_l )
	}.
	\qquad
	\label{eigen:3}
\end{eqnarray}
Next we turn to (\ref{U_j chi}).
With the aid of (\ref{U_j in y}) and (\ref{useful formula})
the LHS of (\ref{U_j chi}) becomes
\begin{eqnarray}
&&	e^{\pi i
		\sum_{l=1}^m
		(1/q_j)
		( L^{2l-1}_{\;2j-1} \nu_{l} y^{2l} 
		- L^{2l}_{\;2j-1} \nu_{l} y^{2l-1} )
	}
	\chi_{k,r,d} ( y^i - L^i_{\;2j-1} /q_j )
	\nonumber \\
& = &	e^{2 \pi i
		\sum_{i=1}^m 
			\{
				- \beta_{2i-1} L^{2i-1}_{\;2j-1} / q_j
				- (1/2) 
				(L^{-1})^{2j}_{\;2i} L^{2i}_{\;2j-1} / q_j
			\}
	} \,
	e^{- 2 \pi i 
		\sum_{i,l=1}^{n-2m}
		d_i ( L^{-1} )^{2m+i}_{\;2m+l} L^{2m+l}_{\;2j-1} /q_j 
	} 
	\nonumber \\ &&
	e^{ 2 \pi i
	\sum_{i=1}^m 
		\{ 
			( k_{2i-1} + \beta_{2i-1} ) y^{2i-1} 
			+ k_{2i-1} \beta_{2i}/\nu_i 
			- (1/2) \nu_i y^{2i-1} y^{2i}
		\}
	}
	\nonumber \\ && 
	e^{ 2 \pi i
		\sum_{i,l=1}^{n-2m}
		d_i ( L^{-1} )^{2m+i}_{\;2m+l} y^{2m+l} 
	}
	\nonumber \\ &&
	\sum_{\sigma_1,\sigma_2,\dots,\sigma_m = -\infty}^\infty
	c_{0,r,d,\sigma} \,
	e^{-2 \pi i
		\sum_{i,l=1}^m 
		(q_i \sigma_i + r_i - \gamma_i) 
		(L^{-1})^{2i}_{\; 2l} L^{2l}_{\;2j-1}/q_j 
	} 
	\nonumber \\ &&
	e^{ 2 \pi i
		\sum_{i,l=1}^m 
		(q_i \sigma_i + r_i + \delta_{ij} - \gamma_i) 
		(L^{-1})^{2i}_{\; 2l} 
		( y^{2l} - k_{2l-1} / \nu_l )
	}
	\label{LHS U_j chi:1}
\end{eqnarray}
To make this coincide with the RHS of (\ref{U_j chi}) 
we have a recursive equation
\begin{eqnarray}
&&	e^{2 \pi i
		\sum_{i=1}^m 
			\{
				- \beta_{2i-1} L^{2i-1}_{\;2j-1}/ q_j
				- (1/2) 
				(L^{-1})^{2j}_{\;2i} L^{2i}_{\;2j-1} / q_j
			\}
	} \,
	e^{- 2 \pi i 
		\sum_{i,l=1}^{n-2m}
		d_i ( L^{-1} )^{2m+i}_{\;2m+l} L^{2m+l}_{\;2j-1} /q_j 
	} 
	\nonumber \\ &&
	e^{-2 \pi i
		\sum_{i,l=1}^m 
		(q_i \sigma_i + r_i - \gamma_i) 
		(L^{-1})^{2i}_{\; 2l} L^{2l}_{\;2j-1}/q_j 
	} \,
	c_{0,r,d,\sigma} 
=	c_{0,(r+1),d,\sigma}.
	\label{equation from U_j chi:1}
\end{eqnarray}
Here $ (r+1) $ means $ ( r_i + \delta_{ij} ) $.
If we define an $ m \times m $ matrix
\begin{equation}
	Y_{ij} 
	= \sum_{l=1}^m (L^{-1})^{2i}_{\; 2l} L^{2l}_{\;2j-1} / q_j 
	= \sum_{l=1}^m \nu_l L^{2l-1}_{\;2i-1} L^{2l}_{\;2j-1}/ ( q_i q_j ),
\end{equation}
it is symmetric as
\begin{equation}
	Y_{ij} - Y_{ji}
	= \sum_{l=1}^m 
		(
		L^{2l-1}_{\;2i-1} \nu_l L^{2l}_{\;2j-1} -
		L^{2l}_{\;2i-1}   \nu_l L^{2l-1}_{\;2j-1} 
		) / ( q_i q_j )
	= \phi_{2i-1,2j-1} / ( q_i q_j )
	= 0
\end{equation}
by virtue of (\ref{magnetic matrix}).
Then the solution of (\ref{equation from U_j chi:1}) is
\begin{eqnarray}
	c_{0,r,d,\sigma} 
& = &
	c_{0,0,d,\sigma} \,
	e^{-2 \pi i
		\sum_{i,j=1}^m 
		\beta_{2i-1} L^{2i-1}_{\;2j-1} r_j / q_j
	} 
	\nonumber \\ &&
	e^{- 2 \pi i 
		\sum_{j=1}^m
		\sum_{i,l=1}^{n-2m}
		d_i ( L^{-1} )^{2m+i}_{\;2m+l} L^{2m+l}_{\;2j-1} r_j / q_j 
	} 
	\nonumber \\ &&
	e^{-2 \pi i
		\sum_{i,j,l=1}^m 
		(q_i \sigma_i + (1/2) r_i - \gamma_i) 
		(L^{-1})^{2i}_{\; 2l} L^{2l}_{\;2j-1} r_j / q_j 
	}.
	\label{c00dn}
\end{eqnarray}
Therefore, (\ref{eigen:3}) becomes
\begin{eqnarray}
	\chi_{k,r,d} ( y^1, y^2, \dots, y^n ) 
& = &
	e^{ 2 \pi i
	\sum_{j=1}^m 
		\{ 
			( k_{2j-1} + \beta_{2j-1} ) y^{2j-1} 
			+ k_{2j-1} \beta_{2j} / \nu_j 
			- (1/2) \nu_j y^{2j-1} y^{2j} 
		\}
	}
	\nonumber \\ && 
	e^{ 2 \pi i
		\sum_{i,l=1}^{n-2m}
		d_i ( L^{-1} )^{2m+i}_{\;2m+l} 
		( 
			y^{2m+l} 
			- \sum_{j=1}^m L^{2m+l}_{\;2j-1} r_j / q_j 
		)
	}
	\nonumber \\ &&
	e^{-2 \pi i
		\sum_{i,j=1}^m 
		\beta_{2i-1} L^{2i-1}_{\;2j-1} r_j / q_j
	} 
	\,
	e^{ \pi i
		\sum_{i,j,l=1}^m 
		\nu_l
		L^{2l-1}_{\;2i-1}
		L^{2l}_{\;2j-1} 
		( r_i / q_i )
		( r_j / q_j )
	} 
	\nonumber \\ &&
	\sum_{\sigma_1,\sigma_2,\dots,\sigma_m = -\infty}^\infty
	c_{0,0,d,\sigma} 
	\nonumber \\ &&
	e^{ 2 \pi i
		\sum_{j,l=1}^m 
		(q_j \sigma_j + r_j - \gamma_j) (L^{-1})^{2j}_{\; 2l} 
		( 
			y^{2l} - k_{2l-1} / \nu_l 
			- \sum_{i=1}^m L^{2l}_{\;2i-1} r_i / q_i 
		)
	}.
	\label{eigen:4}
\end{eqnarray}
Since $ ( U_j )^{q_j} = 1 $ by (\ref{relation1}),
the substitution $ r_j \mapsto r_j + q_j $ must leave $ \chi_{k,r,d} $ invariant.
This substitution gives
\begin{eqnarray}
	\chi_{k,(r+q),d} ( \vect{y} ) 
& = &	e^{ 2 \pi i
	\sum_{i=1}^m 
		\{ 
			( k_{2i-1} + \beta_{2i-1} ) y^{2i-1} 
			+ k_{2i-1} \beta_{2i} / \nu_i 
			- (1/2) \nu_i y^{2i-1} y^{2i} 
		\}
	}
	\nonumber \\ && 
	e^{ 2 \pi i
		\sum_{i,l=1}^{n-2m}
		d_i ( L^{-1} )^{2m+i}_{\;2m+l} 
		( 
			y^{2m+l} 
			- \sum_{p=1}^{m} L^{2m+l}_{\;2p-1} 
			r_p / q_p 
		)
	}
	\nonumber \\ &&
	e^{-2 \pi i
		\sum_{i,l=1}^m 
		\beta_{2i-1} L^{2i-1}_{\;2l-1} 
		r_l / q_l
	} 
	\,
	e^{ \pi i
		\sum_{i,l,p=1}^m 
		\nu_l
		L^{2l-1}_{\;2i-1}
		L^{2l}_{\;2p-1} 
		( r_i / q_i )
		( r_p / q_p )
	} 
	\nonumber \\ && 
	e^{-2 \pi i
		\sum_{i,l=1}^{n-2m}
		d_i ( L^{-1} )^{2m+i}_{\;2m+l} 
		L^{2m+l}_{\;2j-1} 
	}
	\,
	e^{-2 \pi i
		\sum_{i=1}^m 
		\beta_{2i-1} L^{2i-1}_{\;2j-1} 
	} 
	\nonumber \\ &&
	e^{ \pi i
		\{
		\sum_{i,l=1}^m 
		\nu_l L^{2l-1}_{\;2i-1} L^{2l}_{\;2j-1} (r_i/q_i)
		+
		\sum_{l,p=1}^m 
		\nu_l L^{2l-1}_{\;2j-1} L^{2l}_{\;2p-1} (r_p/q_p)
		+
		\sum_{l=1}^m 
		\nu_l L^{2l-1}_{\;2j-1} L^{2l}_{\;2j-1} 
		\}
	} 
	\nonumber \\ &&
	\sum_{\sigma_1,\sigma_2,\dots,\sigma_m = -\infty}^\infty
	c_{0,0,d,(\sigma-1)} 
	\nonumber \\ &&
	e^{-2 \pi i
		\sum_{i,l=1}^m 
		(q_i \sigma_i + r_i - \gamma_i) 
		(1/q_i) \nu_l L^{2l-1}_{\; 2i-1} 
		L^{2l}_{\;2j-1} 
	} 
	\nonumber \\ &&
	e^{ 2 \pi i
		\sum_{i,l=1}^m 
		(q_i \sigma_i + r_i - \gamma_i) 
		(L^{-1})^{2i}_{\; 2l} 
		( 
			y^{2l} - k_{2l-1} / \nu_l 
			- \sum_{p=1}^m L^{2l}_{\;2p-1} 
			r_p / q_p 
		)
	}.
	\label{replaced eigen:4}
\end{eqnarray}
Here $ (\sigma-1) $ is an abbreviation of $ (\sigma_i - \delta_{ij} ) $.
We used (\ref{useful formula}).
Then we have another recursive equation
\begin{eqnarray}
	c_{0,0,d,\sigma}
& = &
	c_{0,0,d,(\sigma-1)} \,
	e^{ -2 \pi i
		\sum_{i,l=1}^{n-2m}
		d_i ( L^{-1} )^{2m+i}_{\;2m+l} L^{2m+l}_{\;2j-1} 
	}
	\, e^{ -2 \pi i
		\sum_{i=1}^m 
		\beta_{2i-1} L^{2i-1}_{\;2j-1} 
	} 
	\nonumber \\ &&
	e^{ \pi i
		\{
		\sum_{i,l=1}^m 
		\nu_l L^{2l-1}_{\; 2i-1} L^{2l}_{\;2j-1} ( r_i/q_i )
	+	\sum_{p,l=1}^m 
		\nu_l L^{2l-1}_{\; 2j-1} L^{2l}_{\;2p-1} ( r_p/q_p )
	+ 	\sum_{l=1}^m 
		\nu_l L^{2l-1}_{\; 2j-1} L^{2l}_{\;2j-1} 
		\}
	} 
	\nonumber \\ &&
	e^{ -2 \pi i
		\sum_{i,l=1}^m 
		( q_i \sigma_i + r_i - \gamma_i) 
		(1/q_i) \nu_l L^{2l-1}_{\; 2i-1} 
		L^{2l}_{\;2j-1} 
	}.
	\label{c00dn-1}
\end{eqnarray}
Remember that
$
	q_i q_j Y_{ij}
	=	\sum_{l=1}^m 
		\nu_l L^{2l-1}_{\; 2i-1} L^{2l}_{\;2j-1} 
	= q_j q_i Y_{ji}
$ is symmetric.
The solution of (\ref{c00dn-1}) is
\begin{eqnarray}
	c_{0,0,d,\sigma}
& = &
	c_{0,0,d,0} 
	\, e^{ -2 \pi i
		\sum_{i,l=1}^{n-2m} \sum_{j=1}^m
		d_i ( L^{-1} )^{2m+i}_{\;2m+l} L^{2m+l}_{\;2j-1} \sigma_j
	}
	\, e^{ -2 \pi i
		\sum_{i,j=1}^m 
		\beta_{2i-1} L^{2i-1}_{\;2j-1} \sigma_j
	} 
	\nonumber \\ &&
	e^{ \pi i
		\{
		\sum_{i,l,j=1}^m 
		\nu_l L^{2l-1}_{\; 2i-1} L^{2l}_{\;2j-1} ( r_i/q_i ) \sigma_j
	+	\sum_{p,l,j=1}^m 
		\nu_l L^{2l-1}_{\; 2j-1} L^{2l}_{\;2p-1} \sigma_j ( r_p/q_p )
		\}
	} 
	\nonumber \\ &&
	e^{ -2 \pi i
		\sum_{i,l,j=1}^m 
		( (1/2) q_i \sigma_i + r_i - \gamma_i) 
		(1/q_i) \, \nu_l L^{2l-1}_{\; 2i-1} 
		L^{2l}_{\;2j-1} \sigma_j
	}.
	\label{c00d0}
\end{eqnarray}
Substituting it into (\ref{eigen:4})
and using (\ref{useful formula}), we get
\begin{eqnarray}
	\chi_{k,r,d} ( \vect{y} ) 
& = &
	e^{- \pi i
		\sum_{j=1}^m \nu_j y^{2j-1} y^{2j} 
	}
	\nonumber \\ &&
	\sum_{\sigma_1,\sigma_2,\dots,\sigma_m = -\infty}^\infty
	c_{0,0,d,0} 
	\nonumber \\ &&
	e^{ \pi i
		\sum_{i,j,l=1}^m 
		\nu_l
		L^{2l-1}_{\;2i-1}
		L^{2l}_{\;2j-1} 
		( q_i \sigma_i + r_i ) 
		( q_j \sigma_j + r_j ) 
		/ (q_i q_j )
	} 
	\nonumber \\ &&
	e^{ 2 \pi i
		\sum_{j=1}^m 
		\beta_{2j-1} 
		\{ y^{2j-1} 
		- \sum_{i=1}^m L^{2j-1}_{\;2i-1} (q_i \sigma_i + r_i) / q_i \}
	} 
	\nonumber \\ &&
	e^{ 2 \pi i
		\sum_{j,l=1}^m 
		(q_j \sigma_j + r_j - \gamma_j) (L^{-1})^{2j}_{\; 2l} 
		\{
			y^{2l} 
			- \sum_{i=1}^m L^{2l}_{\;2i-1} 
			( q_i \sigma_i + r_i) / q_i 
		\}
	} 
	\nonumber \\ && 
	e^{ 2 \pi i
		\sum_{i,l=1}^{n-2m}
		d_i ( L^{-1} )^{2m+i}_{\;2m+l} 
		\{
			y^{2m+l} 
			- \sum_{j=1}^m L^{2m+l}_{\;2j-1} 
			( q_j \sigma_j + r_j ) / q_j 
		\}
	}
	\nonumber \\ &&
	e^{ 2 \pi i
		\sum_{j=1}^m 
		k_{2j-1} 
		\{
			y^{2j-1} + \beta_{2j} / \nu_j 
			- \sum_{l=1}^m 
			L^{2j-1}_{\;2l-1} 
			(q_l \sigma_l + r_l - \gamma_l) / q_l
		\}
	}.
	\label{eigen:5}
\end{eqnarray}
Finally, we are going to solve (\ref{T^l chi}).
Its LHS becomes
\begin{eqnarray}
	e^{2 \pi i \, t^{2m+l}} \chi_{k,r,d}
& = &	
	e^{- \pi i
		\sum_{j=1}^m \nu_j y^{2j-1} y^{2j} 
	}
	\nonumber \\ &&
	\sum_{\sigma_1,\sigma_2,\dots,\sigma_m = -\infty}^\infty
	c_{0,0,d,0} 
	\nonumber \\ &&
	e^{ \pi i
		\sum_{i,j,p=1}^m 
		\nu_p
		L^{2p-1}_{\;2i-1}
		L^{2p}_{\;2j-1} 
		( q_i \sigma_i + r_i ) 
		( q_j \sigma_j + r_j ) 
		/ (q_i q_j )
	} 
	\nonumber \\ &&
	e^{ 2 \pi i
		\sum_{j=1}^m 
		\beta_{2j-1} 
		\{ y^{2j-1} 
		- \sum_{i=1}^m L^{2j-1}_{\;2i-1} (q_i \sigma_i + r_i) / q_i \}
	} 
	\nonumber \\ &&
	e^{ 2 \pi i
		\sum_{j,p=1}^m 
		(q_j \sigma_j + r_j - \gamma_j
		+ \sum_{i=1}^m (L^{-1})^{2m+l}_{\;2i} L^{2i}_{\;2j}
		) 
		(L^{-1})^{2j}_{\; 2p} 
		\{
			y^{2p} 
			- \sum_{i=1}^m L^{2p}_{\;2i-1} 
			( q_i \sigma_i + r_i) / q_i 
		\}
	} 
	\nonumber \\ &&
	e^{ 2 \pi i
		\sum_{p,i=1}^m 
		(L^{-1})^{2m+l}_{\;2p}
		L^{2p}_{\;2i-1} 
		( q_i \sigma_i + r_i) / q_i 
	} 
	\nonumber \\ && 
	e^{ 2 \pi i
		\sum_{i,p=1}^{n-2m}
		( d_i + \delta_i^l ) ( L^{-1} )^{2m+i}_{\;2m+p} 
		\{
			y^{2m+p} 
			- \sum_{j=1}^m L^{2m+p}_{\;2j-1} 
			( q_j \sigma_j + r_j ) / q_j 
		\}
	}
	\nonumber \\ && 
	e^{ 2 \pi i
		\sum_{p=1}^{n-2m}
		\sum_{j=1}^m 
		( L^{-1} )^{2m+l}_{\;2m+p} 
		L^{2m+p}_{\;2j-1} 
		( q_j \sigma_j + r_j ) / q_j 
	}
	\nonumber \\ &&
	e^{ 2 \pi i
		\sum_{j=1}^m 
		( k_{2j-1} + (L^{-1})^{2m+l}_{\;2j-1} )
		\{
			y^{2j-1} + \beta_{2j} / \nu_j 
			- \sum_{p=1}^m 
			L^{2j-1}_{\;2p-1} 
			(q_p \sigma_p + r_p - \gamma_p) / q_p
		\}
	}
	\nonumber \\ &&
	e^{-2 \pi i
		\sum_{j=1}^m 
		(L^{-1})^{2m+l}_{\;2j-1} 
		\{
			\beta_{2j} / \nu_j 
			- \sum_{p=1}^m 
			L^{2j-1}_{\;2p-1} 
			(q_p \sigma_p + r_p - \gamma_p) / q_p
		\}
	}.
	\label{would be chi d+1}
\end{eqnarray}
We put $ \Delta^l k_i = (L^{-1})^{2m+l}_{\; i} $ as before.
The change $ d_p \mapsto d_p + \delta_{p}^{\;l} $ causes 
a change of $ \gamma_j $ as
\begin{eqnarray}
	\Delta^l \gamma_j 
& = &
	\sum_{i=1}^{n-2m}
	( L^{-1} )^{2m+l}_{\;2m+i} L^{2m+i}_{\; 2j}
	\nonumber \\
& = &
	\sum_{i=1}^{n} ( L^{-1} )^{2m+l}_{\;i} L^{i}_{\; 2j}
	- \sum_{i=1}^{m} ( L^{-1} )^{2m+l}_{\;2i-1} L^{2i-1}_{\; 2j}
	- \sum_{i=1}^{m} ( L^{-1} )^{2m+l}_{\;2i}   L^{2i}_{\; 2j}
	\nonumber \\
& = &
	0 - 0
	- \sum_{i=1}^{m} ( L^{-1} )^{2m+l}_{\;2i}   L^{2i}_{\; 2j}
	\label{Delta gamma}
\end{eqnarray}
via (\ref{gamma}) with (\ref{zeros in L}).
Moreover, we can see that
\begin{eqnarray}
	\sum_{p=1}^m
	L^{2j-1}_{\;2p-1} \Delta^l \gamma_p / q_p
& = &
	- \sum_{p,i=1}^m
	L^{2j-1}_{\;2p-1} 
	( L^{-1} )^{2m+l}_{\;2i}   L^{2i}_{\; 2p} / q_p
	\nonumber \\
& = &
	- \sum_{p,i=1}^m
	( L^{-1} )^{2p}_{\;2j} 
	( L^{-1} )^{2m+l}_{\;2i} L^{2i}_{\; 2p} / \nu_j
	\nonumber \\
& = &
	- \sum_{i=1}^m
	( L^{-1} )^{2m+l}_{\;2i} 
	\delta^{2i}_{\;2j} / \nu_j
	\nonumber \\
& = &
	- ( L^{-1} )^{2m+l}_{\;2j} / \nu_j
	\nonumber \\
& = &
	- \Delta^l k_{2j} / \nu_j.
	\label{L gamma}
\end{eqnarray}
Therefore (\ref{would be chi d+1}) becomes
\begin{eqnarray}
	e^{2 \pi i \, t^{2m+l}} \chi_{k,r,d}
& = &	
	e^{2 \pi i
		\sum_{j=1}^m 
		( k_{2j-1} + \Delta k_{2j-1} )
		\Delta k_{2j} / \nu_j
	}
	\nonumber \\ &&
	e^{-2 \pi i
		\sum_{j=1}^m 
		(L^{-1})^{2m+l}_{\;2j-1} 
		( \beta_{2j} / \nu_j 
		+ \sum_{p=1}^m 
		L^{2j-1}_{\;2p-1} \gamma_p / q_p
		)
	}
	\nonumber \\ &&
	e^{- \pi i
		\sum_{j=1}^m \nu_j y^{2j-1} y^{2j} 
	}
	\nonumber \\ &&
	\sum_{\sigma_1,\sigma_2,\dots,\sigma_m = -\infty}^\infty
	c_{0,0,d,0} 
	\nonumber \\ &&
	e^{ \pi i
		\sum_{i,j,p=1}^m 
		\nu_p
		L^{2p-1}_{\;2i-1}
		L^{2p}_{\;2j-1} 
		( q_i \sigma_i + r_i ) 
		( q_j \sigma_j + r_j ) 
		/ ( q_i q_j )
	} 
	\nonumber \\ &&
	e^{ 2 \pi i
		\sum_{j=1}^m 
		\beta_{2j-1} 
		\{ y^{2j-1} 
		- \sum_{i=1}^m L^{2j-1}_{\;2i-1} (q_i \sigma_i + r_i) / q_i \}
	} 
	\nonumber \\ &&
	e^{ 2 \pi i
		\sum_{j,p=1}^m 
		(q_j \sigma_j + r_j - \gamma_j - \Delta \gamma_j
		) 
		(L^{-1})^{2j}_{\; 2p} 
		\{
			y^{2p} 
			- \sum_{i=1}^m L^{2p}_{\;2i-1} 
			( q_i \sigma_i + r_i) / q_i 
		\}
	} 
	\nonumber \\ && 
	e^{ 2 \pi i
		\sum_{i,p=1}^{n-2m}
		( d_i + \delta_i^l ) ( L^{-1} )^{2m+i}_{\;2m+p} 
		\{
			y^{2m+p} 
			- \sum_{j=1}^m L^{2m+p}_{\;2j-1} 
			( q_j \sigma_j + r_j ) / q_j 
		\}
	}
	\nonumber \\ &&
	e^{ 2 \pi i
		\sum_{j=1}^m 
		( k_{2j-1} + \Delta k_{2j-1} )
		\{
			y^{2j-1} + \beta_{2j} / \nu_j 
			- \sum_{p=1}^m 
			L^{2j-1}_{\;2p-1} 
			(q_p \sigma_p + r_p - \gamma_p - \Delta \gamma_p) / q_p
		\}
	}.
	\qquad
\end{eqnarray}
In the course of calculation we used the fact
$
	\sum_{p=1}^n (L^{-1})^{2m+l}_{\;p} L^{p}_{\;2i-1} = 0
$.
With the aid of (\ref{useful formula}), (\ref{Delta gamma})
and the definition (\ref{Z}) we can deduce that
\begin{equation}
	\sum_{j,p=1}^m 
	(L^{-1})^{2m+l}_{\;2j-1} 
	L^{2j-1}_{\;2p-1} \gamma_p / q_p
	= 
	- \sum_{i=1}^{n-2m} Z^{li} \, d_i.
\end{equation}
Thus we reach
\begin{eqnarray}
	e^{2 \pi i \, t^{2m+l}} \chi_{k,r,d}
& = &	
	e^{-2 \pi i
		\sum_{j=1}^m 
		(L^{-1})^{2m+l}_{\;2j-1} \beta_{2j} / \nu_j 
	}
	\, c_{0,0,d,0} \,
	c_{0,0,(d+1),0}^{-1} \,
	\nonumber \\ &&
	e^{2 \pi i
		\{
		\sum_{j=1}^m 
		( k_{2j-1} + \Delta k_{2j-1} )
		\Delta k_{2j} / \nu_j
		+
		\sum_{i=1}^{n-2m} Z^{li} d_i
		\}
	}
	\chi_{k + \Delta k,r,(d+1)}.
\end{eqnarray}
To satisfy (\ref{T^l chi}) we meet another recursive equation
\begin{equation}
	c_{0,0,(d+1),0}
= 
	e^{-2 \pi i
		\sum_{j=1}^m 
		(L^{-1})^{2m+l}_{\;2j-1} \beta_{2j} / \nu_j 
	}
	c_{0,0,d,0}
\end{equation}
and we get the solution
\begin{equation}
	c_{0,0,d,0}
= 
	e^{-2 \pi i
		\sum_{j=1}^m 
		\sum_{l=1}^{n-2m}
		d_l (L^{-1})^{2m+l}_{\;2j-1} \beta_{2j} / \nu_j 
	}
	c_{0,0,0,0}.
	\label{c0000}
\end{equation}
Substituting it into (\ref{eigen:5}) we reach the final result
\begin{eqnarray}
	\chi_{k,r,d} ( \vect{y} ) 
& = &
	c_{0,0,0,0} 
	\, e^{-2 \pi i
		\sum_{j=1}^m 
		\sum_{l=1}^{n-2m}
		d_l (L^{-1})^{2m+l}_{\;2j-1} \beta_{2j} / \nu_j 
	}
	\nonumber \\ &&
	e^{- \pi i
		\sum_{j=1}^m \nu_j y^{2j-1} y^{2j} 
	}
	\nonumber \\ &&
	\sum_{\sigma_1,\sigma_2,\dots,\sigma_m = -\infty}^\infty
	\nonumber \\ &&
	e^{ \pi i
		\sum_{i,j,l=1}^m 
		\nu_l
		L^{2l-1}_{\;2i-1}
		L^{2l}_{\;2j-1} 
		( q_i \sigma_i + r_i ) 
		( q_j \sigma_j + r_j ) 
		/ (q_i q_j )
	} 
	\nonumber \\ &&
	e^{ 2 \pi i
		\sum_{j=1}^m 
		\beta_{2j-1} 
		\{ y^{2j-1} 
		- \sum_{i=1}^m L^{2j-1}_{\;2i-1} (q_i \sigma_i + r_i) / q_i \}
	} 
	\nonumber \\ &&
	e^{ 2 \pi i
		\sum_{j,l=1}^m 
		(q_j \sigma_j + r_j - \gamma_j) (L^{-1})^{2j}_{\; 2l} 
		\{
			y^{2l} 
			- \sum_{i=1}^m L^{2l}_{\;2i-1} 
			( q_i \sigma_i + r_i) / q_i 
		\}
	} 
	\nonumber \\ && 
	e^{ 2 \pi i
		\sum_{k,l=1}^{n-2m}
		d_k ( L^{-1} )^{2m+k}_{\;2m+l} 
		\{
			y^{2m+l} 
			- \sum_{i=1}^m L^{2m+l}_{\;2i-1} 
			( q_i \sigma_i + r_i ) / q_i 
		\}
	}
	\nonumber \\ &&
	e^{ 2 \pi i
		\sum_{j=1}^m 
		k_{2j-1} 
		\{
			y^{2j-1} + \beta_{2j} / \nu_j 
			- \sum_{i=1}^m 
			L^{2j-1}_{\;2i-1} 
			(q_i \sigma_i + r_i - \gamma_i) / q_i
		\}
	}.
	\label{eigen:6}
\end{eqnarray}
This is the result (\ref{chi}) announced previously.
The eigenfunction $ \chi_{k,r,d} $ is determined up to 
a unique normalization constant $ c_{0,0,0,0} $.
Thus we conclude that 
the space of twisted periodic functions over the torus
is an irreducible representation space
of the magnetic algebra $ {\cal A} $.
Finally, we have proved that the set of functions $ \{ \chi_{k,r,d} \} $ 
is a complete orthonormal set in the space of twisted periodic functions
over the torus as announced at (\ref{CONS}).

\section{Eigenfunctions of the magnetic Laplacian}
In this section
we will write down explicitly
solutions of the eigenvalue problem of the magnetic Laplacian
(\ref{Laplacian}) or (\ref{Laplacian in P}),
which is expressed in the $ y $-coordinate as
\begin{eqnarray}
	\Delta f
& = &
	\sum_{j=1}^{m}
	\left[
		\left(
			\frac{\partial}{\partial y^{2j-1}} 
			+ \pi i \nu_j y^{2j} 
			- 2 \pi i \beta_{2j-1}
		\right)^2 f
		+
		\left(
			\frac{\partial}{\partial y^{2j}} 
			- \pi i \nu_j y^{2j-1} 
			- 2 \pi i \beta_{2j}
		\right)^2 f
	\right]
	\nonumber \\
&&	+
	\sum_{k=1}^{n-2m}
	\left(
		\frac{\partial}{\partial y^{2m+k}} 
		- 2 \pi i \beta_{2m+k}
	\right)^2 f.
	\label{Laplacian in y}
\end{eqnarray}
The function $ f $ must satisfy the twisted periodic condition
(\ref{twist condition}).
We will obtain solutions using the Fourier analysis
that is developed in the previous section.

Here we would like to describe the outline of our method.
As mentioned in the previous section,
the Laplacian commutes with the magnetic shift operators,
$ \{ U_j, V_j, W_k \} $.
Hence the labels 
$ (r,d) = ( r_1, r_2, \dots, r_m, d_1, d_2, \dots, d_{n-2m} ) $,
which are defined in
(\ref{label r})-(\ref{represent W}),
are `good' quantum numbers.
Moreover, the Laplacian commutes with the longitudinal momentum operators
$ \{ P_{2m+k} \} $.
Hence the corresponding momentum eigenvalues $ \{ k_{2m+k} \} $
are also good quantum numbers
and are related to the labels $ d_l $ via (\ref{k:d}).
On the other hand,
the Laplacian does not commute with the transverse momentum operators
$ \{ P_{2j-1}, P_{2j} \} $.
Hence the transverse momentum eigenvalues $ \{ k_{2j-1} \} $
do not remain good quantum numbers.
The Laplacian admits a new set of good quantum numbers
$ ( n_1, n_2, \cdots, n_m ) $,
which will be introduced soon later.
It will be revealed that
eigenfunctions of the Laplacian are actually
matrix elements of a unitary transformation,
\begin{equation}
	\psi_{n,r,d} (k_1,k_3,\dots,k_{2m-1}) 
	= \bra k,r,d | n,r,d \ket,
\end{equation}
which relates the quantum numbers $ n $'s to $ k $'s.
In the $ k $-space
it is rather easy to get eigenfunctions 
by the standard method of a harmonic oscillator.
On the other hand, the set of eigenfunctions 
of the momenta and magnetic shifts,
\begin{equation}
	\chi_{k,r,d} (y^1,y^2,\dots,y^n) 
	= \bra y | k,r,d \ket,
\end{equation}
plays a role a unitary transformation
which bridges between the momentum space and the real space
like the usual Fourier transformation.
Hence 
the Laplacian eigenfunctions are transformed 
into the $ y $-coordinate representations by
\begin{equation}
	\psi_{n,r,d} (y^1,y^2,\dots,y^n) 
	= \bra y | n,r,d \ket
	= 
	\int_{-\infty}^{\infty} d k_1 d k_3 \cdots d k_{2m-1}
	\bra y | k,r,d \ket \, \bra k,r,d | n,r,d \ket.
	\label{transform:k to y}
\end{equation}
This will give the desired result.

Now let us carry out the program outlined above.
We define creation and annihilation operators
associated with the transverse momenta as
\begin{equation}
	a_j^\dagger 
	= \frac{1}{\sqrt{4 \pi \nu_j}} ( P_{2j-1} -i P_{2j} ),
	\quad
	a_j 
	= \frac{1}{\sqrt{4 \pi \nu_j}} ( P_{2j-1} +i P_{2j} )
	\qquad ( j=1, \dots, m ).
	\label{creation annhilation}
\end{equation}
It is easily verified that $ [ a_j, a_k^\dagger ] = \delta_{jk} $.
Then the Laplacian (\ref{Laplacian in P}) becomes
\begin{equation}
	- \Delta =
	\sum_{j=1}^m 4 \pi \nu_j 
	\left( a_j^\dagger a_j + \frac{1}{2} \right)
	+
	\sum_{k=1}^{n-2m} ( P_{2m+k} )^2.
	\label{Laplacian in a}
\end{equation}
The eigenstate $ | {\mit\Omega} \ket $ for the lowest eigenvalue 
satisfies
\begin{eqnarray}
	0 
& = &	\bra k | a_j | {\mit\Omega} \ket
	= \frac{1}{\sqrt{4 \pi \nu_j}} 
	\bra k | ( P_{2j-1} +i P_{2j} ) | {\mit\Omega} \ket
	\nonumber \\ 
& = &	\frac{1}{\sqrt{4 \pi \nu_j}} 
	\left( 
		2 \pi k_{2j-1} + \nu_j \frac{\partial}{\partial k_{2j-1}}
	\right)
	\bra k | {\mit\Omega} \ket.
\end{eqnarray}
Here we used 
(\ref{represent P 2j-1}) and (\ref{represent P 2j}).
The solution is 
\begin{equation}
	\bra k | {\mit\Omega} \ket 
	= e^{ - \pi \sum_{j=1}^m (k_{2j-1})^2 / \nu_j }.
\end{equation}
States for higher eigenvalues
are generated by creation operators as
\begin{eqnarray}
	\bra k | n \ket
& = &
	\frac{1}{\sqrt{ n_1 ! \cdots n_m ! }}
	\bra k | 
	(a_1^\dagger)^{n_1} 
	\cdots
	(a_m^\dagger)^{n_m}
	| {\mit\Omega} \ket
	\nonumber \\
& = &
	\frac{1}{\sqrt{ n_1 ! \cdots n_m ! }}
	\, \prod_{j=1}^m
	\left[ 
		\frac{1}{\sqrt{4 \pi \nu_j}} 
		\bigg( 
		2 \pi k_{2j-1} - \nu_j \frac{\partial}{\partial k_{2j-1}}
		\bigg)
	\right]^{n_j}
	\bra k | {\mit\Omega} \ket
	\nonumber \\
& = &
	\frac{1}{\sqrt{ n_1 ! \cdots n_m ! }}
	\, e^{ \pi \sum_{j=1}^m (k_{2j-1})^2 / \nu_j } 
	\, \prod_{j=1}^m
	\left[ 
		\frac{1}{\sqrt{4 \pi \nu_j}} 
		\bigg( 
		- \nu_j \frac{\partial}{\partial k_{2j-1}}
		\bigg)
	\right]^{n_j}
	e^{ - 2 \pi \sum_{j=1}^m (k_{2j-1})^2 / \nu_j }
	\nonumber \\ &&
	\label{energy eigen in k}
\end{eqnarray}
for $ n_1, n_2, \dots, n_m = 0, 1, 2, \dots $.
We are suppressing other labels $ (r,d) $.
In the appendix B we prove that 
\begin{eqnarray}
	\int_{-\infty}^{\infty} dk \,
	e^{2 \pi i k z}
	\cdot
	e^{   \pi k^2 / \nu} 
	\left( - \nu \frac{\partial}{\partial k} \right)^n
	e^{-2 \pi k^2 / \nu}
& = &
	\sqrt{\nu} \, e^{ \pi \nu z^2 } 
	\left( - i \frac{\partial}{\partial z} \right)^n
	e^{- 2 \pi \nu z^2 } 
	\nonumber \\
& = &
	\sqrt{\nu} \left( i \sqrt{2 \pi \nu} \right)^n
	e^{- \pi \nu z^2 } \,
	H_{n} \Big( z \sqrt{2 \pi \nu} \, \Big).
	\quad
	\label{tiny formula}
\end{eqnarray}
In the second line $ H_n( \xi ) $ is the $ n $-th Hermite polynomial.
Substituting
(\ref{chi}) and (\ref{energy eigen in k}) into
(\ref{transform:k to y})
and applying (\ref{tiny formula}) we obtain
\begin{eqnarray}
	\bra y | n,r,d \ket
& = &
	\int_{-\infty}^{\infty} 
	dk_1 dk_3 \cdots d_{2m-1} \,
	\bra y | k,r,d \ket
	\bra k,r,d | n,r,d \ket
	\nonumber \\
& = &
	c
	\, e^{-2 \pi i
		\sum_{j=1}^m 
		\sum_{l=1}^{n-2m}
		d_l (L^{-1})^{2m+l}_{\;2j-1} \beta_{2j} / \nu_j 
	}
	\nonumber \\ &&
	e^{- \pi i
		\sum_{j=1}^m \nu_j y^{2j-1} y^{2j} 
	}
	\nonumber \\ &&
	\sum_{\sigma_1,\sigma_2,\dots,\sigma_m = -\infty}^\infty
	\nonumber \\ &&
	e^{ \pi i
		\sum_{i,j,l=1}^m 
		\nu_l
		L^{2l-1}_{\;2i-1}
		L^{2l}_{\;2j-1} 
		( q_i \sigma_i + r_i ) 
		( q_j \sigma_j + r_j ) 
		/ (q_i q_j )
	} 
	\nonumber \\ &&
	e^{ 2 \pi i
		\sum_{j=1}^m 
		\beta_{2j-1} 
		\{ y^{2j-1} 
		- \sum_{i=1}^m L^{2j-1}_{\;2i-1} (q_i \sigma_i + r_i) / q_i \}
	} 
	\nonumber \\ &&
	e^{ 2 \pi i
		\sum_{j,l=1}^m 
		(q_j \sigma_j + r_j - \gamma_j) (L^{-1})^{2j}_{\; 2l} 
		\{
			y^{2l} 
			- \sum_{i=1}^m L^{2l}_{\;2i-1} 
			( q_i \sigma_i + r_i) / q_i 
		\}
	} 
	\nonumber \\ && 
	e^{ 2 \pi i
		\sum_{k,l=1}^{n-2m}
		d_k ( L^{-1} )^{2m+k}_{\;2m+l} 
		\{
			y^{2m+l} 
			- \sum_{i=1}^m L^{2m+l}_{\;2i-1} 
			( q_i \sigma_i + r_i ) / q_i 
		\}
	}
	\nonumber \\ &&
	\prod_{j=1}^m
	\left[
		\sqrt{ \frac{\nu_j}{n_j !} }
		\Bigg(
		\frac{i}{\sqrt{2}} 
		\Bigg)^{n_j}
		e^{ - \pi \nu_j 
		\{
			y^{2j-1} + \beta_{2j} / \nu_j 
			- \sum_{i=1}^m 
			L^{2j-1}_{\;2i-1} 
			(q_i \sigma_i + r_i - \gamma_i) / q_i
		\}^2 
	}
	\right.
	\nonumber \\ &&
	\qquad \left.
	H_{n_j}
	\bigg(
		\sqrt{2 \pi \nu_j}
		\Big\{
			y^{2j-1} + \beta_{2j} / \nu_j 
			- \sum_{i=1}^m 
			L^{2j-1}_{\;2i-1} 
			(q_i \sigma_i + r_i - \gamma_i) / q_i
		\Big\}
	\bigg)
	\right].
	\quad
	\label{solution}
\end{eqnarray}
This is the main result of this paper.
Using (\ref{k:d}),
we can calculate
eigenvalues of the Laplacian (\ref{Laplacian in a}) as
\begin{equation}
	- \frac{1}{2} \Delta \psi_{n,r,d}
	=
	\left[
		\,
		\sum_{j=1}^m
		2 \pi \nu_j \bigg( n_j + \frac{1}{2} \bigg)
		+
		\frac{1}{2}
		\sum_{k=1}^{n-2m} 
		\bigg\{
			\sum_{l=1}^{n-2m} 
			2 \pi ( d_l - \alpha_{2m+l} )
			(L^{-1})^{2m+l}_{\;2m+k}
		\bigg\}^2
	\right]
	\psi_{n,r,d}.
	\label{energy eigenvalue}
\end{equation}
Eigenvalues depend on
quantum numbers
$ n_1, n_2, \dots, n_m = 0, 1, 2, \dots $
and
$ d_1, d_2, \dots, d_{n-2m} $ $ = 0, \pm 1, \pm 2, \dots $
but not on
$ r = ( r_1, r_2, \dots, r_m ) \in 
\Z_{q_1} \times \Z_{q_2} \times \cdots \times \Z_{q_m} $.
Thus 
each eigenvalue is degenerated by $ q_1 q_2 \cdots q_m $ folds
as predicted in (\ref{dim}).
If a ratio $ \nu_i / \nu_j (i \ne j) $ is rational,
degeneracy happens more.
On the other hand,
if $ \alpha_{2m+l} = 0 $,
the eigenvalue for $ -d_l $ coincides with the one for $ d_l $.
If $ \alpha_{2m+l} = 1/2 $,
the eigenvalue for $ (-d_l + 1) $ coincides with the one for $ d_l $.
Moreover,
for specific values of 
$ \{ (L^{-1})^{2m+l}_{\;2m+k} \} $
we may meet more multi-fold degeneracy.

Let us discuss physical meanings of the eigenvalue (\ref{energy eigenvalue}).
It is energy of an electrically-charged particle
moving in the magnetic field in the torus.
In (\ref{energy eigenvalue})
we put the coefficient $ -1/2 $ in front of $ \Delta $
to adjust the equation to the conventional Schr{\"o}dinger equation.
In the context of classical mechanics,
the particle exhibits a cyclic motion 
with the frequency $ \nu_j $ in the $ (y^{2j-1}, y^{2j}) $-plane
for each $ j = 1, 2, \dots, m $.
And it exhibits a uniform straight motion
along the $ y^{2m+k} $-axis.
The whole motion is a superposition of those cyclic and straight motions.
When we turn to quantum mechanics,
energy of the cyclic motion is quantized and results in
the so-called Landau level
$ 2 \pi \nu_j ( n_j + 1/2 ) $.
On the other hand, the longitudinal momentum $ P_{2m+k} $
associated with the straight motion is quantized to be
$
	2 \pi k_{2m+k}
	=
	2 \pi
	\sum_{l=1}^{n-2m} 
	( d_l - \alpha_{2m+l} ) (L^{-1})^{2m+l}_{\;2m+k}
	=
	2 \pi 
	\sum_{l=1}^{n-2m} 
	d_l (L^{-1})^{2m+l}_{\;2m+k}
	- \beta_{2m+k}
$
with integers
$ ( d_1, d_2, \dots, d_{n-2m} ) $
as explained in (\ref{k:d}).
Along the course of the straight motion 
the particle flies around the torus
and picks up the so-called Aharonov-Bohm effect.
Then the momentum is shifted by the Aharonov-Bohm parameters
$ ( \beta_{2m+1}, \dots, \beta_n ) $.
Accordingly, the kinetic energy of the straight motion is also quantized.
The total energy is then given as (\ref{energy eigenvalue}).

Moreover, let us examine meanings of other Aharonov-Bohm parameters
$ ( \beta_1, \dots, \beta_{2m} ) $.
These do not affect the energy (\ref{energy eigenvalue})
and hence they have a geometric significance 
rather than a physical significance.
To understand their meaning
we rewrite the eigenfunction (\ref{eigen:6}) as
\begin{eqnarray}
	\chi_{k,r,d} ( \vect{y} ) 
& = &
	c 
	\,
	e^{-2 \pi i
		\sum_{j,i=1}^m 
		\beta_{2j-1} 
		L^{2j-1}_{\;2i-1} 
		\gamma_i / q_i 
	} 
	\,
	e^{-2 \pi i
		\sum_{j=1}^m 
		\sum_{l=1}^{n-2m}
		d_l (L^{-1})^{2m+l}_{\;2j-1} \beta_{2j} / \nu_j 
	}
	\nonumber \\ &&
	e^{- \pi i
		\sum_{j=1}^m \nu_j y^{2j-1} y^{2j} 
	}
	\nonumber \\ &&
	e^{ 2 \pi i
		\sum_{j=1}^m \beta_{2j-1} y^{2j-1} 
	} 
	\nonumber \\ &&
	\sum_{\sigma_1,\sigma_2,\dots,\sigma_m = -\infty}^\infty
	\nonumber \\ &&
	e^{ \pi i
		\sum_{i,j,l=1}^m 
		\nu_l
		L^{2l-1}_{\;2i-1}
		L^{2l}_{\;2j-1} 
		( q_i \sigma_i + r_i ) 
		( q_j \sigma_j + r_j ) 
		/ (q_i q_j )
	} 
	\nonumber \\ &&
	e^{ 2 \pi i
		\sum_{j,l=1}^m 
		(q_j \sigma_j + r_j - \gamma_j) (L^{-1})^{2j}_{\; 2l} 
		\{
			y^{2l} - \beta_{2l-1} / \nu_l 
			- \sum_{i=1}^m L^{2l}_{\;2i-1} 
			( q_i \sigma_i + r_i ) / q_i 
		\}
	} 
	\nonumber \\ && 
	e^{ 2 \pi i
		\sum_{k,l=1}^{n-2m}
		d_k ( L^{-1} )^{2m+k}_{\;2m+l} 
		\{
			y^{2m+l} 
			- \sum_{i=1}^m L^{2m+l}_{\;2i-1} 
			( q_i \sigma_i + r_i ) / q_i 
		\}
	}
	\nonumber \\ &&
	e^{ 2 \pi i
		\sum_{j=1}^m 
		k_{2j-1} 
		\{
			y^{2j-1} + \beta_{2j} / \nu_j 
			- \sum_{i=1}^m 
			L^{2j-1}_{\;2i-1} 
			(q_i \sigma_i + r_i - \gamma_i) / q_i
		\}
	}.
	\label{eigen:7}
\end{eqnarray}
{}From (\ref{Delta gamma}) we have
\begin{equation}
	\sum_{j,i=1}^m 
	\beta_{2j-1} L^{2j-1}_{\;2i-1} \gamma_i / q_i 
	= 
	- 
	\sum_{l=1}^{n-2m}
	\sum_{j=1}^{m} 
	d_l ( L^{-1} )^{2m+l}_{\;2j} 
	\, \beta_{2j-1} / \nu_j.
\end{equation}
Then (\ref{eigen:7}) is rewritten as
\begin{eqnarray}
	\chi_{k,r,d} ( \vect{y} ) 
& = &
	c 
	\,
	e^{2 \pi i
		\sum_{l=1}^{n-2m}
		\sum_{j=1}^{m} 
		d_l 
		\{
			( L^{-1} )^{2m+l}_{\;2j} 
			\beta_{2j-1} 
			-
			( L^{-1} )^{2m+l}_{\;2j-1} 
			\beta_{2j} 
		\} / \nu_j
	}
	\nonumber \\ &&
	e^{- \pi i
		\sum_{j=1}^m \nu_j y^{2j-1} y^{2j} 
	}
	\nonumber \\ &&
	e^{ 2 \pi i
		\sum_{j=1}^m \beta_{2j-1} y^{2j-1} 
	} 
	\nonumber \\ &&
	\sum_{\sigma_1,\sigma_2,\dots,\sigma_m = -\infty}^\infty
	\nonumber \\ &&
	e^{ \pi i
		\sum_{i,j,l=1}^m 
		\nu_l
		L^{2l-1}_{\;2i-1}
		L^{2l}_{\;2j-1} 
		( q_i \sigma_i + r_i ) 
		( q_j \sigma_j + r_j ) 
		/ ( q_i q_j )
	} 
	\nonumber \\ &&
	e^{ 2 \pi i
		\sum_{j,l=1}^m 
		(q_j \sigma_j + r_j - \gamma_j) (L^{-1})^{2j}_{\; 2l} 
		\{
			y^{2l} - \beta_{2l-1} / \nu_l 
			- \sum_{i=1}^m L^{2l}_{\;2i-1} 
			( q_i \sigma_i + r_i ) / q_i 
		\}
	} 
	\nonumber \\ && 
	e^{ 2 \pi i
		\sum_{k,l=1}^{n-2m}
		d_k ( L^{-1} )^{2m+k}_{\;2m+l} 
		\{
			y^{2m+l} 
			- \sum_{i=1}^m L^{2m+l}_{\;2i-1} 
			( q_i \sigma_i + r_i ) / q_i 
		\}
	}
	\nonumber \\ &&
	e^{ 2 \pi i
		\sum_{j=1}^m 
		k_{2j-1} 
		\{
			y^{2j-1} + \beta_{2j} / \nu_j 
			- \sum_{i=1}^m 
			L^{2j-1}_{\;2i-1} 
			(q_i \sigma_i + r_i - \gamma_i) / q_i
		\}
	}.
	\label{eigen:8}
\end{eqnarray}
Thus we can see that the parameters
$ ( \beta_1, \dots, \beta_{2m} ) $
induce a displacement
\begin{equation}
	( y^{2j-1}, y^{2j} )
	 \to 
	 ( y^{2j-1} + \beta_{2j} / \nu_j,
	 y^{2j} - \beta_{2j-1} / \nu_j )
\end{equation}
of the profile $ | \chi_{k,r,d} ( \vect{y} ) | $.
This is the geometric significance of
the transverse Aharonov-Bohm parameters.

\section{Conclusion}
Here we summarize our discussions.
As well-known, a $ U(1) $ gauge field replaces 
the partial derivative by the covariant derivative
and generates a magnetic field.
As a natural extension of the eigenvalue problem
of the ordinary Laplacian in the $ n $-torus,
we formulated the eigenvalue problem of the magnetic Laplacian.
The ordinary Laplacian admits continuous Abelian symmetry
and therefore the usual Fourier analysis is applicable.
However,
the magnetic Laplacian does not admit continuous Abelian symmetry
and therefore the usual Fourier analysis is not applicable to it.
Hence,
we developed an alternative method, 
which became an extension of the usual Fourier analysis.
We identified symmetry structure of the magnetic Laplacian
and defined the magnetic translation group (MTG),
which is discrete and non-Abelian in general.
Moreover, we defined the magnetic algebra 
by extending the MTG.
We proved that the space of functions on which the magnetic Laplacian acts
is an irreducible representation space of the magnetic algebra.
By diagonalizing the maximal Abelian subalgebra of the magnetic algebra
we obtained a complete orthonormal set of functions 
$ \{ \chi_{k,r,d} ( \vect{y} ) \} $
over the magnetic torus;
those functions are labeled by a set of good quantum numbers $ (k,r,d) $.
It was rather easy to diagonalize the magnetic Laplacian 
in the $ \vect{k} $-space representation.
Applying a unitary transformation by $ \{ \chi_{k,r,d} ( \vect{y} ) \} $
to the eigenstate of the magnetic Laplacian,
we finally obtained the eigenfunction 
in the $ \vect{y} $-space representation.
The eigenvalues of the magnetic Laplacian
were naturally interpreted as sums of
energies 
of cyclic motions in the transverse directions to the magnetic field
and
energies
of linear motions in the longitudinal direction to it.

New results of this paper are
the definition and representations of the magnetic algebra,
the proof of irreducibility of the space of twisted periodic functions
as a representation space of the magnetic algebra,
the complete orthogonal set of functions (\ref{chi})
which provides a basis of the extended Fourier analysis,
the eigenfunctions (\ref{solution}) of the magnetic Laplacian
in explicit forms,
and the eigenvalues (\ref{energy eigenvalue}).

Before closing this paper
we would like to discuss briefly possible directions for further development.
We treated only the Laplace operator in this paper 
but for application to physics
it is more desirable to treat the Schr{\"o}dinger operator
\begin{equation}
	H = - \frac{1}{2} \Delta + V,
\end{equation}
which has a potential energy term $ V $.
The potential $ V $ is a periodic function;
in the $ t $-coordinate it satisfies
$ V ( t^1, \dots, t^i + 1, \dots, t^n ) 
= V ( t^1, \dots, t^i,     \dots, t^n ) $
for each $ i $.
It acts on the twisted periodic function $ f( \vect{t} ) $ 
by multiplication.
To take the potential term into account 
we may introduce new operators $ X^k $ by
\begin{equation}
	( X^j f ) ( \vect{t} )
	= 
	e^{2 \pi i \, t^{j} } f ( \vect{t} )
	\qquad
	(j=1, 2, \dots, 2m),
	\label{X}
\end{equation}
which belong to the same family of operators $ T^k $ of (\ref{T}).
Then any periodic potential operator can be expanded as
\begin{eqnarray}
	V ( t^1, \dots, t^n ) 
& = &
	\sum_{\sigma_1, \sigma_2, \dots, \sigma_n = - \infty}^\infty
	c_{\sigma} \,
	e^{2 \pi i ( \sigma_1 t^1 + \cdots + \sigma_n t^n ) }
	\nonumber \\
& = &
	\sum_{\sigma_1, \sigma_2, \dots, \sigma_n = - \infty}^\infty
	c_{\sigma} \,
	( X^1 )^{\sigma_1}
	\cdots
	( X^{2m} )^{\sigma_{2m}}
	( T^1 )^{\sigma_{2m+1}} \cdots
	( T^{n-2m} )^{\sigma_{n}}.
\end{eqnarray}
We can easily calculate commutators of $ X $'s with other operators
to get an algebra which is an extension of the magnetic algebra.
The resulted algebra is isomorphic to
the so-called noncommutative torus \cite{Connes}
although we do not yet examine these relation thoroughly.

Another direction for future development
is to solve an eigenvalue problem of the Dirac operator in the $ n $-torus
in the background magnetic field.
We also construct supersymmetric field theory,
which have both scalar and spinor fields as its constituents
to pursue a new mechanism of supersymmetry breaking.

\section*{Acknowledgments}
S.T. thanks T. Iwai and I. Tsutsui
for their continuous encouragement.
The authors thank the Yukawa Institute for Theoretical Physics 
at Kyoto University. 
Discussions during the 
YITP workshop YITP-W-02-18 on
``Quantum Field Theories: Fundamental Problems and Applications''
were useful to complete this work. 
This work was supported in part 
by the Grant-in-Aid for Scientific Research 
of Ministry of Education, Science, and Culture, Grant No.15540277.


\renewcommand{\thesection}{Appendix A.}
\renewcommand{\theequation}{A.\arabic{equation}}
\section{Distribution of zeros}
As announced at (\ref{zeros in L})
we prove 
existence of a transformation matrix $ L $ 
that has zeros in the pattern
\begin{equation}
	L =
	\left(
	\begin{array}{lll}
	L^{2i-1}_{\;2p-1} & L^{2i-1}_{\;2q} & L^{2i-1}_{\;2m+r} \\
	L^{2j}_{\;2p-1}   & L^{2j}_{\;2q}   & L^{2j}_{\;2m+r}   \\
	L^{2m+k}_{\;2p-1} & L^{2m+k}_{\;2q} & L^{2m+k}_{\;2m+r} 
	\end{array}
	\right)
	=
	\left(
	\begin{array}{lll}
		\ast & 0    & 0   \\
		\ast & \ast & 0   \\
		\ast & \ast & \ast 
	\end{array}
	\right)
	\label{zeros in L:Appendix}
\end{equation}
with $ i,j,p,q = 1, \dots, m $ and $ k,r = 1, \dots, n-2m $.

In $ \R^n $ we have an anti-symmetric bilinear form $ B $.
We say that a vector $ \vect{u} $ is longitudinal with respect to $ B $
if it satisfies
\begin{equation}
	B ( \vect{u}, \vect{v} ) = 0
\end{equation}
for arbitrary $ \vect{v} \in \R^n $.
We call
\begin{equation}
	M^0 =
	\{
		\vect{u} \in \R^n
		\, | \,
		\forall \vect{v} \in \R^n, \,
		B ( \vect{u}, \vect{v} ) = 0
	\}
\end{equation}
a longitudinal vector subspace.
Let $ (t^1,t^2,\dots,t^n) $ be the coordinate system 
that expresses $ B $ in the standard form
\begin{equation}
	B 
	= \frac12 \sum_{i,j=1}^n \phi_{ij} \, dt^i \wedge dt^j
	= \sum_{j=1}^m q_j \, dt^{2j-1} \wedge dt^{2j}.
\end{equation}
as in (\ref{standard form}).
Let $ (y^1,y^2,\dots,y^n) $ be another coordinate system 
that is related to $ (t^1,t^2,\dots,t^n) $
by a linear transformation $ y^i = \sum_{j=1}^n L^i_{\,j} \, t^j $.
The matrix $ L $ is not yet specified.
Basis vectors generated by these coordinates are related as
\begin{eqnarray}
&&	\frac{\partial}{\partial t^j}
	=
	\sum_{i=1}^n
	\frac{\partial y^i}{\partial t^j} \,
	\frac{\partial}{\partial y^i}
	=
	\sum_{i=1}^n
	L^i_{\,j} \,
	\frac{\partial}{\partial y^i},
	\label{t in y}
	\\
&&	\frac{\partial}{\partial y^j}
	=
	\sum_{i=1}^n
	\frac{\partial t^i}{\partial y^j} \,
	\frac{\partial}{\partial t^i}
	=
	\sum_{i=1}^n
	( L^{-1} )^i_{\,j} \,
	\frac{\partial}{\partial t^i}.
	\label{y in t}
\end{eqnarray}
Define vector subspaces $ M^- $ and $ M^+ $ of $ \R^n $ as
\begin{eqnarray}
&&	M^- =
	\R \frac{\partial}{\partial t^{1}} \oplus
	\R \frac{\partial}{\partial t^{3}} \oplus \cdots \oplus
	\R \frac{\partial}{\partial t^{2m-1}},
\\
&&	M^+ =
	\R \frac{\partial}{\partial t^{2}} \oplus
	\R \frac{\partial}{\partial t^{4}} \oplus \cdots \oplus
	\R \frac{\partial}{\partial t^{2m}}.
\end{eqnarray}
Then $ M^- \oplus M^+ \oplus M^0 = \R^n $.
Now let us remember that $ \R^n $ is equipped 
with inner product structure.
Then the two-form $ B : \R^n \times \R^n \to \R $ 
can be regarded as an antisymmetric operator $ \hat{B} : \R^n \to \R^n $.
As a square of a linear operator
$ \hat{B}^2 $ is well-defined and 
becomes a symmetric operator and therefore 
is diagonalizable by an orthogonal transformation.
$ \hat{B}^2 $ has non-positive eigenvalues.
The eigenspace $ W^0 $ associated with the zero eigenvalue of $ \hat{B}^2 $
coincides with $ M^0 $.
Of course,
\begin{equation}
	\left\{
	\frac{\partial}{\partial t^{2m+1}}, 
	\frac{\partial}{\partial t^{2m+2}}, \dots,
	\frac{\partial}{\partial t^{n}}
	\right\}
\end{equation}
is a basis of $ M^0 = W^0 $. We take an orthonormal basis
\begin{equation}
	\left\{
	\frac{\partial}{\partial y^{2m+1}}, 
	\frac{\partial}{\partial y^{2m+2}}, \dots,
	\frac{\partial}{\partial y^{n}}
	\right\}
\end{equation}
of $ W^0 $.
This implies that
\begin{equation}
	L^{2i-1}_{\;2m+r}        = L^{2j}_{\;2m+r} =
	(L^{-1})^{2i-1}_{\;2m+r} = (L^{-1})^{2j}_{\;2m+r} = 0
\end{equation}
in (\ref{t in y}) and (\ref{y in t}).

Let us step into a difficult part of the proof.
Each eigenspace associated with each nonzero eigenvalue of $ \hat{B}^2 $
can be decomposed into two-dimensional subspaces 
such that
each two-dimensional subspace $ W_j $
is irreducible with respect to the action of $ \hat{B} $.
Thus we get an orthogonal decomposition
\begin{equation}
	\R^n = W_1 \perp W_2 \perp \cdots \perp W_m \perp W^0.
	\label{orthogonal decomposition}
\end{equation}
Next we define vector subspaces
\begin{eqnarray}
&&	W_j^+ = W_j \cap ( M^+ \oplus M^0 ),
	\\
&&	W_j^- = W_j \cap ( M^+ \oplus M^0 )^\perp,
\end{eqnarray}
then we can show that
both $ W_j^+ $ and $ W_j^- $ have one dimension.
First, note that 
$ \dim W_j^+ + \dim W_j^- = \dim W_j = 2 $.
If $ \dim W_j^+ = 2 $, $ W_j^+ $ coincides with $ W_j $ itself.
Since the two-form $ B $ is degenerated on $ ( M^+ \oplus M^0 ) $,
it must be degenerated also on 
$ W_j = W_j^+ \subset ( M^+ \oplus M^0 ) $.
This contradicts the fact that $ W_j $
is irreducible with respect to $ \hat{B} $.
On the other hand,
if $ \dim W_j^- = 2 $, $ W_j^- $ coincides with $ W_j $ itself. 
Then 
we can take an arbitrary one-dimensional subspace 
$ M_{-1} \subset W_j = W_j^- \subset ( M^+ \oplus M^0 )^\perp $.
Since $ W_j $ is irreducible with respect to $ \hat{B} $,
$ B $ is degenerated on $ M_{-1} \oplus M^+ \oplus M^0 $.
This contradicts the fact that 
$ M^+ \oplus M^0 $ is a maximal degenerated subspace of $ B $.
Hence we conclude that $ \dim W_j^+ = \dim W_j^- = 1 $.

We take a normalized vector 
$
	\partial / \partial y^{2j}
$
of $ W_j^+ $.
And we take another normalized vector 
$
	\partial / \partial y^{2j-1}
$
of $ W_j^- $ 
such that 
$ \nu_j = B( \partial / \partial y^{2j-1}, \partial / \partial y^{2j} ) > 0 $
for each $ j = 1, \dots, m $.
Then we obtain a complete orthonormal basis
$ \{ \partial / \partial y^i \, | \, i = 1,2, \dots, n \} $
that expresses $ B $ in the standard form
$ B = \sum_{j=1}^m \nu_j \, dy^{2j-1} \wedge dy^{2j} $.

Since
$
	\partial / \partial y^{2q}
	\in W_q^+ 
	\subset  M^+ \oplus M^0
$,
we can say that
\begin{equation}
	(L^{-1})^{2i-1}_{\;2q} = 0
\end{equation}
in (\ref{y in t}).
By an elementary argument of linear algebra we can say that
\begin{equation}
	L^{2i-1}_{\;2q} = 0.
\end{equation}
The proof is over.

\renewcommand{\thesection}{Appendix B.}
\renewcommand{\theequation}{B.\arabic{equation}}
\section{Fourier transformation of the Hermite polynomials}
In our convention the Hermite polynomial is defined as
\begin{equation}
	H_n ( \xi ) 
	= (-1)^n \,
	e^{\xi^2} \frac{d^n}{d \xi^n} e^{-\xi^2}.
	\label{Hermite polynomial}
\end{equation}
The formula (\ref{tiny formula}) can be deduced 
by a partial integration and a change of variables as
\begin{eqnarray}
&&
	\int_{-\infty}^{\infty} dk \,
	e^{2 \pi i k z}
	\cdot
	e^{   \pi k^2 / \nu} 
	\left( - \frac{\partial}{\partial k} \right)^n
	e^{-2 \pi k^2 / \nu}
	\nonumber \\
& = &
	\int_{-\infty}^{\infty} dk \,
	e^{-2 \pi k^2 / \nu}
	\left( \frac{\partial}{\partial k} \right)^n
	e^{2 \pi i k z}
	\cdot
	e^{   \pi k^2 / \nu} 
	\nonumber \\
& = &
	e^{ \pi \nu z^2 } 
	\int_{-\infty}^{\infty} dk \,
	e^{-2 \pi k^2 / \nu}
	\left( \frac{\partial}{\partial k} \right)^n
	e^{   \pi ( k + i \nu z)^2 / \nu} 
	\nonumber \\
& = &
	e^{ \pi \nu z^2 } 
	\int_{-\infty}^{\infty} dk \,
	e^{-2 \pi k^2 / \nu}
	\left( - \frac{i}{\nu} \, \frac{\partial}{\partial z} \right)^n
	e^{   \pi ( k + i \nu z)^2 / \nu} 
	\nonumber \\
& = &
	e^{ \pi \nu z^2 } 
	\left( - \frac{i}{\nu} \, \frac{\partial}{\partial z} \right)^n
	\int_{-\infty}^{\infty} dk \,
	e^{- \pi ( k - i\nu z )^2 / \nu
	   - 2 \pi \nu z^2 } 
	\nonumber \\
& = &
	e^{ \pi \nu z^2 } 
	\left( - \frac{i}{\nu} \, \frac{\partial}{\partial z} \right)^n
	\sqrt{\nu} \,
	e^{- 2 \pi \nu z^2 }.
	\label{tiny appedix}
\end{eqnarray}

\baselineskip 6mm 

\end{document}